\documentclass[final]{IEEEtran}
\usepackage[english]{babel}
\usepackage[utf8]{inputenc}
\usepackage{amsmath,amsfonts}
\usepackage{graphicx}
\usepackage{epstopdf}
\usepackage{caption}
\usepackage{subcaption}
\usepackage[colorinlistoftodos]{todonotes}
\usepackage[noadjust]{cite}
\usepackage{url}
\usepackage{balance}
\usepackage{algorithmic,algorithm}

\title{Route Optimization of Electric Vehicles \\ based on Dynamic Wireless Charging}
\author{\IEEEauthorblockN{
Dimitrios Kosmanos, 
Leandros Maglaras,~\IEEEmembership{Senior Member,~IEEE},
Michalis Mavrovouniotis,~\IEEEmembership{Member,~IEEE},
Sotiris Moschoyiannis,
Antonios Argyriou~\IEEEmembership{Senior Member,~IEEE}, 
Athanasios Maglaras, and Helge Janicke}

\thanks{
Corresponding author: H. Janicke (email: heljanic@dmu.ac.uk).} 

\thanks{D. Kosmanos and Antonios Argyriou are in the Department of Electrical \& Computer Engineering,  University of Thessaly, Volos, Greece}
\thanks{L. Maglaras and Helge Janicke are in the Department of Computing Technology, De Montfort University, Leicester, UK.}
\thanks{M. Mavrovouniotis is with the School of Science and Technology, Nottingham Trent University, Nottingham UK.}
\thanks{S. Moschoyiannis is with the Department of Computer Science, University of Surrey, Guildford UK.}
\thanks{A. Maglaras is with the Department of Electrical Engineering,T.E.I. of Thessaly, Larissa, Greece.}
}

\begin{document}
\maketitle

\begin{abstract}
One of the barriers to adoption of Electric Vehicles (EVs) is the anxiety around the limited driving range. Recent proposals have explored charging EVs on the move, using dynamic wireless charging which enables power exchange between the vehicle and the grid while the vehicle is moving. In this article, we focus on the intelligent routing of EVs in need of charging so that they can make most efficient use of the so-called {\it Mobile Energy Disseminators} (MEDs) which operates as mobile charging stations. We present a method for routing EVs around MEDs on the road network, which is based on constraint logic programming and optimisation using a graph-based shortest path algorithm. The proposed method exploits Inter-Vehicle (IVC) communications in order to eco-route electric vehicles. 
We argue that combining modern communications between vehicles and state of the art technologies on energy transfer, the driving range of EVs can be extended without the need for larger batteries or overtly costly infrastructure. We present extensive simulations in city conditions that show the driving range and consequently the overall travel time of electric vehicles is improved with intelligent routing in the presence of MEDs.
\end{abstract}


\section{Introduction}
\label{sec:intro}


There is increasing interest among government agencies, research institutions and industry around the globe in improving urban living while reducing the environmental impact. The term {\it smart city} has been coined to describe the city of tomorrow in which modern intelligent technologies, such as IT communication systems, sensors, machine learning, data analytics, come together to provide better services to the citizens. Just like a complex system, a smart city can monitor, coordinate and manage information, connectivity and assets that citizens need every day and adapt to accommodate their demands. One of the basic components of this environment is envisaged to be the next generation of vehicles that combine new sensing, communication and social capabilities. By providing mobile wireless sensing and communications, vehicles can facilitate data access, which is fundamental to realising the premise of smart cities. 

Smart vehicles are expected to be a part of a Vehicular Ad hoc NETwork (VANET), a mobile ad hoc network of cars that has been proposed to enhance traffic safety and provide comfort applications to drivers. A VANET has some unique characteristics such as high mobility of nodes, while cars must follow predefined routes; messages that come from several applications, with different priority levels; high interference, in a noisy environment, and so on.  Using the on-board unit,  vehicles can communicate with each other as well as with road side units (RSUs) enabling smart application solutions but also enhanced road safety and traffic management. According to several works, e.g., see \cite{vegni2013smart}, smart vehicles exhibit five features: self-driving, safety driving, social driving, electric vehicles and mobile applications. In this paper, we focus on electric vehicles. 

One of the prohibiting factors for the adoption of the Electric Vehicles (EVs) across Europe is the driving range \cite{thiel2015electric}, \cite{vtc2017}. That is, the range the vehicle can cover before it needs to be re-charged.
The lack of supporting charging infrastructure is a pivotal prohibiting factor. The deployment of charging infrastructure is a hard problem \cite{charging2015infrastructure} as it inadvertently requires changes to the existing civil infrastructure and these are costly and take a long time to implement. The car industry is experimenting with larger and more powerful batteries - new Tesla and WV EVs have been released with powerful batteries that promise to cover up to 400km  without intermediate charge. However, it is argued that in the future batteries of reduced capacity should be used, mainly for environmental reasons.

It transpires there is a need for new approaches to charging electric vehicles that overcome the lack of supporting infrastructure and the difficulty of adapting the existing civil infrastructure, i.e., road network, without requiring new batteries that take up most space in the car and are not environmentally friendly. {/it Dynamic wireless charging} is a technology that is still in the R\&D phase. A number of companies are actively developing dynamic wireless charging solutions, both in the research and testing phases. BMW  has already demonstrated wireless charging with the i8 model. Tesla motors also has already produced the Plugless Model S that can use wireless inductive charging at home.  Wireless charging can be the key enabler for electric vehicles if they are to surpass the convenience of gas cars \cite{covic2013modern}. Preliminary analysis, e.g., see \cite{plugless}, suggests that even the most far-out ideas around wireless charging may become reality sooner than most expect . 

\begin{figure}[!ht]
\centering
\includegraphics[width=0.40\textwidth]{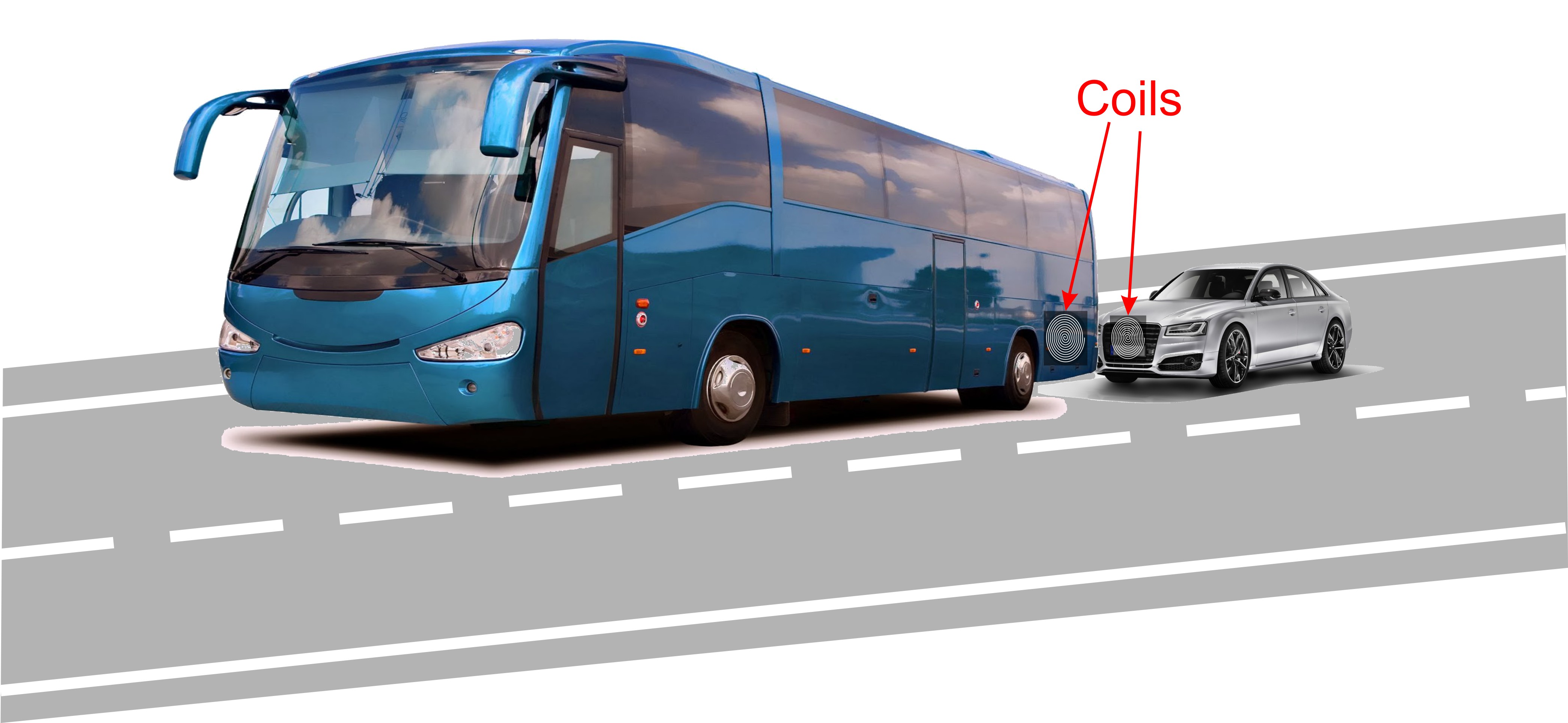}
\caption{Wireless charging of EV using spiral coils}
\label{Figure1}
\end{figure}

This drives the investigation towards integrated solutions that allow EVs to charge on the move. In \cite{maglaras2014cooperative, maglaras2015-IJACSA} the authors have proposed a novel idea for increasing the driving range without requiring a significant change in existing road infrastructure. The idea builds on deploying buses and HGVs, LGVs or trucks, as mobile charging stations, the so-called {\it Mobile Energy Disseminators} (MEDs)  \cite{maglaras2014cooperative}.  While a bus is moving along its normal route an EV in need of charging attaches itself to it and charges via wireless power transmission, as shown in Figure \ref{Figure1}. 

Buses (inner city) and trucks (motorways) repeatedly move at prescribed routes that are scheduled well in advance. Hence, the EV can meet them by appointment at specific locations. The process is similar to charging of aircraft in flight. When the bus finishes its round trip it will return to the fixed static charging station where it will either fully charge or change the batteries. 

In this paper, the focus is on describing the mechanics of the proposed dynamic wireless charging of EVs and the challenges that arise. An EV in need of charge would typically have a choice of MEDs to which it could attach itself. The main contribution of this paper concerns the intelligent routing of EVs in need of charging, and more specifically a solution that draws upon constraint logic programming (CLP) and a graph-based shortest path algorithm (cf Section \ref{problem_formulation}). The optimisation problem of (re)routing is considered under a range of criteria and priorities.

Extensive simulations were conducted in city conditions in order to evaluate the proposed "on the move" charging technique (cf Section \ref{evalaute}). With different initial energy conditions for all the EVs of the simulation, two different charging systems are compared: one uses a static charging station (SCS) only and the second combines a SCS with a MED. The experiments show that the driving range and consequently the overall travel time is improved by about four times in the dynamic charging system involving MEDs.

The remainder of the paper is organised as follows. Section \ref{sec:rel-work} discusses related work and places the research within that of the wider community. Section \ref{MEDs} introduces the key concepts and the overall architecture of the proposed system. Section \ref{problem_formulation} presents the problem formulation of routing electronic vehicles given the presence of static and mobile stations.  Section \ref{evalaute} presents simulation parameters, describes the evaluation of the method and discusses economic benefits of the proposed method. Section \ref{concl} concludes the article.

\section{Related Work}
\label{sec:rel-work}

The wireless power transmission technology is being applied for a number of years now in many areas of electrical appliances, like speakers, music and sound transmission generally, alarm systems, electric bells, and electrical  facilities  of low power in general. In the field of wireless charging of electric vehicles, there are many architectures and special experimental systems that have already been proposed, built and implemented (e.g. Korea reports \cite{jung2012high}). In some of  these infrastructures the locations (points) used for charging are either fixed (static stations) installed either under the surface of streets and in other public locations (i.e. garages) or on lightning columns \cite{telewatt_2013}, \cite{greencar} on the road side. The wireless power transmission in the proposed system is achieved using the Tesla coil method, with spiral coils installed on the vehicles.  

Previous work on charging electric vehicles mainly focuses on static charging stations \cite{xu2013optimal},  swappable batteries \cite{adler2014online}, eco-routing of vehicles \cite{de2016intention} or dynamic charging \cite{lukic2013cutting} that is based on static sources. Based on the work in \cite{lukic2013cutting} dynamic wireless charging of vehicles promises to partially or completely eliminate the overnight charging of electric vehicles through the use of dynamic chargers that may be installed on the roads to keep the vehicle batteries continuously charged, thus making electric vehicles more attractive. The use of dynamic wireless charging  may increase driving range and reduce the size of the battery pack of an electric vehicle. On the other hand, this leads to increased safety concerns and infrastructure costs.
 
Previous work on dynamic wireless charging has not considered the solution of moving energy charging stations that can charge vehicles, which are also on the move, in order to reduce the range anxiety and increase the reliability of EVs. Authors in \cite{telewatt_2013} presented a solution called {\it Telewatt} that involves the reuse of existing public lighting infrastructure for vehicle charging. It does so by exploiting the excessive power of the lamps mostly at night. This system that supports wireless charging between the infrastructure and the moving vehicles  raises health issues related to the leaking magnetic flux. In another work in \cite{ning2013compact} authors present a system that can charge vehicles through inductive coupling. The prototype for EV that was developed at Oak Ridge National Laboratory (ORNL) in the United States achieved efficiency of nearly 90\% for 3 kW power delivery.  However, systems that are based on inductive coupling between the grid and a moving car can cause power pulsations in the vehicle battery and the grid supply. This can result in deterioration on the battery service life of EVs as well as a drop on the power quality of the grid \cite{miller2014demonstrating}.

The disadvantages of these methods can be summarised as follows. 
\begin{itemize}
 \item Charging an EV from a stationery charger introduces a large or small delay due to 
	\begin{enumerate}
    \item the change of the route of the movement of the EV to the loading point (location), 
    \item the need of parking for a sufficient period of time to charge, and 	     \item the restoration of the EV at the initial route. 
   \end{enumerate}
 \item The infrastructure would need to be extensive and consequently expensive \cite{nagendra2014detection}  
 \item The (energy transfer efficiency) performance of the charging method would be relatively insufficient (or low) due to the inherent operational difficulties of the systems (e.g., distance, parallelism, etc.)
\end{itemize}

The solution we propose in this paper builds on the use of inner city buses as MEDs, hence it does not suffer from the pitfalls associated with static charging stations. In addition, it uses buses or trucks for the dynamic charging, so predefined moving charging stations which have predefined scheduled routes along the existing road network, rather than vehicle-to-vehicle (V2V) charging schemes that have been discussed in the literature \cite{v2v2012charging}.


The EVs attach themselves to one or more MEDs during some part of their journey and until they have enough energy to reach their destination (or get to the closest static charging station). In this way,  electric cars are charged “on the fly” and their range is increased while moving along the road. Hence, our proposal does not require significant changes to the existing road network and civil infrastructure \cite{machiels2013design,5289760,telewatt_2013} and, unlike other proposals \cite{Guho_2013}, does not pose any health 
hazards.

\section{Dynamic charging and Mobile Energy Disseminators}\label{MEDs}

The dynamic wireless charging system is based on the combination of vehicular communications and inductive  power transfer (IPT) among the energy carriers and the electric vehicles. IPT allows efficient and real-time energy exchange
where the vehicles involved can play an active role in the procedure.

\subsection{Energy transfer via IPT}
\label{energy-transfer}

Using the IPT wireless method, a 10-minute charge would provide a driver with an energy charging of 3 - 8 kWh of electric energy, which is equivalent to about 9 - 23 miles travel distance. The United States fuel economy estimates that 35 kWh equals 100 miles. The energy charging 3 - 8 kWh requires 20 - 50 kW charging rate from the moving charging stations (see Table \ref{table1}). This travel distance corresponds to 30 - 78 percent of the drivers average daily travel distance. In real-world terms, that means typical urban American drivers could cover 78 percent of their average daily travel of 23 miles on a 10-minute charge with charging rate 50 kW. European drivers fare even better; a 10-minute charge with charging rate 50 kW under this wireless scenario would cover nearly two days of a typical European’s driving habits, which amounts to about 20 kilometers or 12.5 miles per day \cite{pasaoglu2012driving}. 
\begin{center}
\begin{table} [h]
\centering
\caption{Miles per 10-minute charge for electric cars \cite{brinknews} \\
\small{* This is for a 30 amp public charging station.}}
\label{arr:param}
 \begin{tabular}{||c c||} 
 \hline
 Method & Value\\ [0.5ex]
 \hline\hline
 Tesla Supercharger & 56.7  \\ 
 \hline
 Mobile Energy Disseminator (MED) & 22.85  \\
 \hline
 Public Charging Station* & 3.7  \\
  \hline
\end{tabular}
\label{table1}
\end{table}
\end{center}
In the case that the charging rate would be 20 kW, a 10-minute charge would cover about 9 miles or about 15 kilometers. By comparison, a public 30 amp wired charging station provides electric cars with just 3.7 miles of range on a 10-minute charge; it takes about an hour at a typical public wired charging station to provide just 22 miles of range to an electric car.

\subsection{EVs and MEDs in a VANET architecture}
\label{vanets-arch}

The use of mobile nodes as relay nodes  is common in vehicular ad hoc networks (VANETs). In a VANET, mobile  nodes can serve as carriers or disseminators of useful information ~\cite{rehman2016adaptive}. Defining influential spreaders, nodes that can disseminate the information to a large part of the
network effectively, is an open issue in ad hoc networks \cite{basaras2013detecting}.
In VANETs, nodes with predefined or repeating routes that can cover a wide range of a city region can play the role of roadside units in terms of message dissemination. By exploiting their mobility these disseminating nodes can provide even  higher quality-of-service (QoS).

\begin{figure}
\begin{center}
\includegraphics[width=0.5\textwidth]{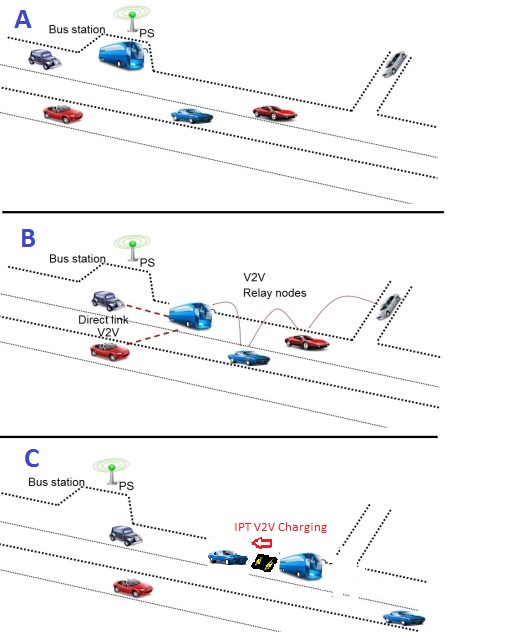}
\caption {Application example of a Mobile Energy Disseminator: In A, Contactless Wireless-Consistency charging is used to deliver charging to a bus; in B, V2V communication between MED and EV; in C, EV re-charges from the bus using IPT.}
\label{fig-ivc}
\end{center}
\end{figure}

Following a similar approach the proposed dynamic wireless charging system is using special nodes,  buses or trucks, that act as energy sources to EVs that are in energy need. The architecture of the proposed system is shown in Figure \ref{fig-ivc}. These vehicles, which are called MEDs, use electric plug in connection or IPT in order to refill starving EVs.  Buses can play the role of MEDs in urban environments, since they follow predefined scheduled routes and their paths cover a major part  of a city, while trucks can play the role of energy chargers mainly on highways. Buses can be fully charged when parked, before beginning their scheduled trip, and can be continuously charged along their journey by IPT stations installed at bus stops (See Figure \ref{fig-ivc}). 
Vehicles that book charging places on the same MED can create clusters/platoons where the MED will play the role of the clusterhead \cite{santini2017consensus}.

The buses or trucks (MEDs) run on electric power. They will have battery systems for their movement, which are used exclusively by the bus or truck (MED). At the same time they carry other systems of special batteries with more energy, which will only be used for charging of EV vehicles in motion. The energy of these batteries will be able to cover the energy need of several  EVs.  The total energy of the charging batteries of the bus is expected to be greater than 200 kWh. The energy of the batteries of an ordinary EV is about 50 kWh, hence the amount of the energy of the batteries carried by the bus will be capable to serve 4 EVs for a total recharge and more for partial recharges.  The charging rate will be 20-50 kW (cf. Section \ref{arr:param}), so that the required charging voltage shall be relatively low. Finally, the bus or truck will carry the mechanisms necessary for the connection and transfer of energy from the MED to the EVs.
The EV charging process will be as follows.


\begin{enumerate}
\item EV contacts MED and makes an appointment (time, location).
\item  EV drives near the MED and creates a platoon with it to initiate the charging process. 
\item The MED charges the EV via a loose connection device consisting of 2 coils, of plain form or better of spiral conical form for greater efficiency and ease of connection. These coils can be of different diameter and perhaps even of different numbers of turns. Another solution comprises using 2 coils, one of which with larger diameter which is mounted on the EV and the other with the smaller diameter on the MED. E / M Shielding will be available  on all 2 vehicles. The 2 coils will be properly covered, and will be uncovered during the charging process. 


\item Vehicles come close and using wireless communication, like an advanced cruise control system, controlled for safety reasons by the MED and  while in motion, come in such a position that the smaller diameter’s MED coil comes close enough to or enters in the EV coil. 
\end{enumerate}


In this article we investigate the wireless power transfer via a loose connection device consisting of 2 coils, of plain or conical form. Due to the design of the coils, and the very close position between them, the energy transfer efficiency will be more than 90\%. 
The proposed design has a similar functionality with the typical Tesla coils \cite{stark2004wireless}, or Rogowski Coil \cite{ramboz1996machinable}. With this solution, electric power is transferred to the vehicle through an electrically generated magnetic field.  The basic functionality of the charging process is comparable to charging via a cable.  An innovative, induction-based mechanism that is developed by Siemens eCar Powertrain Systems \cite{haq2010new} can be also used to offer a significantly higher degree of convenience, when compared with charging via a cable.

A major concern when dealing with strong magnetic
fields, such as those used in wireless power
transfer, has to do with the impact on living organisms. By only
turning on the coils when a compatible electric
vehicle is over the primary charging pad, the charging
system eliminates the possibility that a person or
animal could be affected by the strong fields created.
Another issue with safety has to do with the presence of
metal objects at or close to primary charging pads.
These objects can cause hazardous conditions and
can interfere with WPT. To address this problem,
a foreign object detection system can be deployed in future 
to determine when objects are on top of the primary
coils. In such situations the system will not energise
the transmitting coil so as to avoid damage to the
vehicle and/or charging system.



\subsection{Communication among entities}
\label{communication}

To state its presence each MED or SCS periodically broadcasts cooperative awareness messages (CAM). Each beacon message consists of a node identifier (Vid), node location, scheduled trip (a subset of set L), current charging capability (CC) and energy value (E=KWh), and the queue time at SCS or waiting time (wt) at MED appointment point. CC is the current energy that the mobile charging station can afford to dispose of to charge the vehicle without jeopardising its own needs. These messages are disseminated by all vehicles that effectively act as relay nodes.

\section{Routing EVs in need of charging}
\label{problem_formulation}

\subsection{Problem Formulation: Constrained Shortest Path}
The problem of routing EVs can be presented using a directed 
weighted graph. Let $G=(N,A)$ be a weighted graph where $N$ is a set of points, e.g., 
road intersections or static charging stations (SCS) and $A=\{(i,j)\mid i,j \in N, i \neq j\}$ 
is a set of arcs (links) connecting two points. SCSs are defined as 
$S = \{s_0, \ldots, s_v\}$ and a set of dummy nodes that represent possible multiple visits 
to the same static recharging station is defined as $S' = \{s_{m+1}, \ldots, s_{m+h}\}$ such 
that $S \cup S' \subseteq N$. Each SCS $i$ is associated with a 
waiting time $wt_i$. 

An EV can also receive energy by MEDs
that visit a predefined cyclic route of MED points $M = \{m_0, \ldots, m_u\}$. 
Similarly with the SCS, a set of dummy nodes may represent possible 
multiple visit to the same MED point defined as $M'$ such that $M \cup M' \subseteq N$.
An EV can attach to a MED at any point in its route and start charging. 
Note that the charging rate of MED is always higher than the 
consumption rate. Similar with the SCS, each MED point $i$
has a waiting time $wt_i$ 
This is because an 
EV may need to wait to a point until a MED is available or arrives. MEDs and SCSs 
accept/reject demands of EVs in an intelligent way, i.e., to minimize the route of the 
vehicles at the best possible way or to distribute energy at the best possible 
way (defined by the communication system)

Each arc $(i,j) \in A$ is associated with a non-negative travel time
$dt_{ij} \in \mathbb{R}^+$ and a non-negative energy needed to travel 
$c_{ij} \in \mathbb{R}^+$ when points $i$ and $j$ are connected otherwise 
$dt_{ij}=c_{ij}=\infty$. The weight matrix of the problem is defined as 
$\mathbf{D}=\{dt_{ij}\}_{n \times n}$. 

The objective of the problem is to route a $K$ set of EVs in the best
possible way, i.e., minimum travel time. The problem can be formulated as 
a multiple constrained shortest path problem. Every $k$th EV has a battery of 
$Q^k$ capacity, starting point $s^k$ and destination point $e^k$. The travel time 
is defined by the driving ($dt$), the charging ($ct$) and waiting times ($wt$) at 
different SCS or MED points (if needed). The energy level at point $i$ is 
defined as $\epsilon^k_i$. Hence, the initial energy 
level is defined as $\epsilon^k_s$. 

Let $x_{ij}^k$ and $y_{ij}^k$ be 
binary decision variables that define whether EV $k$ 
passed from point $i$ to $j$ and whether EV $k$ received energy 
from a MED from point $i$ to $j$, respectively. Also, let $z_i^k$ and $q_i^k$ 
be binary decision variables that defines the SCS where EV $k$ received 
energy and the MED point where EV $k$ attached with a MED, respectively. 
All variables used in this paper are summarised in Table~\ref{tab1}.

\begin{table*}[!t]
\begin{center}
\caption{Mathematical symbols used in this paper.}
\footnotesize
\label{tab1}
\begin{tabular}{||l l||}\hline\hline
Symbol & Description \\\hline\hline
$G=(N,A)$ & Weighted graph \\\hline
$N$ & Set of points \\\hline
$A$ & Set of links\\\hline
$(i,j)$ & Link between points $i$ and $j$\\\hline
$dt_{ij}$ & Drive time between points $i$ and $j$\\\hline
$c_{ij}$ & Energy consumed between points $i$ and $j$\\\hline
$\rho$ & Induced energy\\\hline
$S$ & set of static recharging stations\\\hline
$S'$ & extension of $S$ that represent multiple visits \\\hline
$M$ & set of MED points \\\hline
$M'$ & extension of $M$ that represent multiple visits\\\hline
$SW_i(t)$ & SCS $i$ waiting time at period $t$ \\\hline
$MW_i(t)$ & MED point $i$ waiting time at period $t$\\\hline
$k$ & EV id \\\hline
$K$ & set of EVs\\\hline
$s^k$ & start point of an EV \\\hline
$e^k$ & destination point of an EV\\\hline
$Q^k$ & energy capacity of EV $k$ \\\hline
$\epsilon_i^k$ & energy level of EV $k$ at point $i$\\\hline
$x^k_{ij}$ & binary decision variable to identify the route of EV $k$\\\hline
$y^k_{ij}$ & binary decision variable to identify path where EV $k$ received energy from a MED\\\hline
$z_i^k$ & binary decision variable to identify if an EV $k$ received energy from SCS $i$ \\\hline
$q_i^k$ & binary decision variable to identify the MED point an EV $k$ has attached to receive mobile energy \\\hline
$ct_i$ & charging time at SCS $i$\\\hline
$wt_i$ & waiting time at SCS $i$ or MED point $i$\\\hline
$v$ & number of SCSs \\\hline
$u$ & number of MED points \\\hline
\hline\hline
\end{tabular}
\end{center}
\end{table*}

The objective to minimize the travel time of EVs is given next:
\begin{equation}
\begin{aligned}
\min \sum_{k \in K} (\sum_{(i,j) \in A, i \neq j} (dt_{ij}x_{ij}^k) + 
\sum_{i \in S \cup S'}(ct_{i}+wt_{i})z_i^k +  \\ \sum_{i \in M \cup M'} (wt_{i}q_i^k))
\label{eq1}
\end{aligned}
\end{equation}

s.t.
\begin{equation}
\sum_{j \in N}x^{k}_{ij} - \sum_{j \in N}x^{k}_{ji} = \begin{cases}
1,~~\mbox{if $i = s^k$};\\
-1,\mbox{if $i= t^k$};\\
0,~~\mbox{otherwise}
\end{cases} , \forall i \in N, \forall k \in K
\label{eq0}
\end{equation}

\begin{equation}
x_{ij}^k - y_{ij}^k \geq 0, \forall k \in K, \forall j \in N, \forall i \in N, i \neq j,
\label{eq2}
\end{equation}

\begin{equation}
\begin{aligned}
\epsilon^k_j \leq \epsilon^k_i - (c_{ij})x_{ij}^k + (\rho_2d_{ij})y_{ij}^kx_{ij}^k + Q^k(1-x_{ij}^k), \\ \forall k \in K, \forall j \in N, \forall i \in N, i \neq j,
\label{eq3}
\end{aligned}
\end{equation}

\begin{equation}
\epsilon^k_i \geq 0,~~ \forall k \in K, \forall i \in N,
\label{eq5.1}
\end{equation}

\begin{equation}
\epsilon^k_i \leq Q^k, ~~\forall k \in K, \forall i \in N,
\label{eq5.2}
\end{equation}

\begin{equation}
\epsilon^k_i = Q^kz_i^k,~~ \forall k \in K, \forall i \in S \cup S',
\label{eq8}
\end{equation}

\begin{equation}
\epsilon^k_i \geq c_{ij}, \forall k \in K, \forall i \in N, \exists j \in S \cup S' \cup M \cup M', i \neq j,
\label{eq6}
\end{equation}

\begin{equation}
x_{ij}^k,y_{ij}^k \in \{0,1\}, ~\forall k \in K, \forall i \in N, \forall j \in N, i \neq j,
\label{eq7}
\end{equation}

\begin{equation}
z^k_i \in \{0,1\}, \forall k \in K, \forall i \in S \cup S',
\end{equation}

\begin{equation}
q^k_i \in \{0,1\}, \forall k \in K, \forall i \in M \cup M',
\end{equation}
where $ct_{i}$ is the charging time from a charging 
station or visit  $i$ (for a MED the charging time is already 
embedded to the tour in Equation~(\ref{eq3}), 
and $wt_{i}$ is the waiting time at charging station (or a MED's point) $i$. 

Constraint (\ref{eq0}) ensures flow conservation of the route;
constraint (\ref{eq2}) ensures that whenever an EV receives energy 
from a MED while moving always consumes energy;
constraint (\ref{eq3}) ensures that an EV has enough 
energy to move to the next point (including MED's points); 
constraint (\ref{eq5.1}) and (\ref{eq5.2}) 
ensures that energy level never falls under zero or exceeds its 
capacity; constraint (\ref{eq8}) ensures that an EV is fully charged at 
static energy station; constraint (\ref{eq6}) ensures that an EV has enough 
energy to reach at least one recharging static station or MED point.

The feasibility of an EV $k$ route can be identified by the current
energy level and the total energy needed for the route such that
energy must not be negative, as follows: 
\begin{equation}
\epsilon^k_s - \left(\sum_{(i,j) \in A} c_{ij}\right) + \rho \geq 0
\label{eq111}
\end{equation}
where $\epsilon^k_s$ is the initial energy level, $c_{ij}$ the energy consumed from points $i$ to $j$ and 
$\rho$ is the induced energy. 

The key differences of the proposed shortest path problem (described above) with the traditional shortest 
path problem are: \\
a) multiple shortest paths are required, and\\
b) energy constraints are imposed \\
The proposed problem is more challenging and realistic because not all shortest routes are feasible 
due to the energy constraints; see Equation~(\ref{eq111}) and also one shortest route may affect the remaining
shortest routes. For example, if an EV is currently charging at a SCS; then the other EVs will possibly
have to wait (i.e., increasing the queue time of the SCS) or find a shorter route via another SCS.

\subsection{Solution Method}
Since the problem is a shortest path problem it can be solved by 
several existing optimization algorithms efficiently (i.e., in polynomial time). 
In this paper, we consider the well-known Dijkstra's algorithm \cite{Dijkstra1959} to 
calculate the shortest route, e.g., minimize the travel time in Eq~\ref{eq1}, for EV $k$ from 
its starting point $s^k$ to its destination point $e^k$. However, the problem has 
several constraints that need to be addressed and by simply using the Dijkstra's 
algorithm from $s^k$ to $e^k$ may result to an infeasible route, i.e., Equation~(\ref{eq111}) 
does not hold.

The key idea of the proposed solution method is to initially check whether the 
route calculated by Dijkstra's algorithm satisfies Equation~(\ref{eq111}), meaning that
it has sufficient energy to reach the destination. If the the route is 
feasible then the EV should begin its route without any energy recharging 
consideration. Otherwise, it needs to find a point, either static or moving, 
to recharge its battery in order to have sufficient energy to reach the 
destination as shown in Algorithm \ref{alg1}. 

For this case Dijkstra's algorithm is used again to find the best point
to receive energy from. Since there may be several
static charging stations or MED points that the EV can choose, several 
Dijkstra's calculations are performed, one for each point, and the best one 
is selected as shown in Algorithm \ref{alg2}. The criteria to identify the 
best energy point depends on the total travel time, including waiting time, charging
time and driving time. In addition, the energy point selected needs 
to be feasible, i.e., the current energy level of the EV needs to be enough to reach 
the selected energy point (i.e., constraint (\ref{eq6})). Hence, the energy points that 
cannot be reached according to Equation~(\ref{eq111}) are discarded.

\begin{algorithm} [!t]
\caption{FindShortestPath($k$,$s^k$,$e^k$)}
\label{alg1}
\begin{algorithmic}[1] 
\STATE \textbf{INPUT} EV information, e.g., id, source and destination 
\STATE $FinalRoute^k$ $\leftarrow \emptyset$ ~~~~~~~~~~~~\% final route of $k$ EV
\STATE $T^k \leftarrow$ 0 ~~~~~~~~~~~~~~~~~~~~~~~~~\%  travel time of $k$ EV
\STATE $R^k \leftarrow$ Dijkstra($s^k,e^k$)~~~~~~~~~~\% partial route of $k$ EV
\IF {($R^k$ is \textit{feasible})}
\STATE $FinalRoute^k \leftarrow R^k$ 
\ELSE
\STATE $p \leftarrow$ FindBestEnergyPoint($s^k$)
\STATE $FinalRoute^k \leftarrow$ Dijkstra($s^k,p$) $\cup$ Dijkstra($p,e^k$)
\ENDIF
\STATE $T^k \leftarrow Cost(FinalRoute^k)$ 
\STATE \textbf{OUTPUT} $FinalRoute^k$ \% feasible route to travel verified by Equation \ref{eq111}
\STATE \textbf{OUTPUT} $T^k$  \% travel time using Equation~(\ref{eq1}) but for a single EV
\end{algorithmic}
\end{algorithm} 

Finally, when the energy point is selected the shortest path using the Dijkstra's
algorithm is calculated from the selected energy point to the destination. 
Note that in case this path is still not feasible because the energy level may not 
be sufficient to travel from the energy point to the destination the process 
in Algorithm \ref{alg2} can be repeated from the current position, e.g., the 
energy point to the next energy point.

\begin{algorithm} [!t]
\caption{FindBestEnergyPoint($s^k$)}
\label{alg2}
\begin{algorithmic}[1] 
\STATE \textbf{INPUT} current point of EV $k$
\STATE $best \leftarrow \emptyset$ ~~~~~~~\% route of a best energy point
\STATE $p$  ~~~~~~~~~~~~~~~~~\% best energy point
\FOR{($i \in S \cup M$)}
\STATE $R^k \leftarrow$ Dijkstra($s^k,i$) 
\IF{(($Cost(R^k) < Cost(best)$) \&\& ($R^k$ is \emph{feasible}))}
\STATE $p \leftarrow$ $i$ 
\STATE $best \leftarrow R^k$
\ENDIF
\ENDFOR
\STATE \textbf{OUTPUT} $p$  ~~~~~~~\% best energy point
\end{algorithmic}
\end{algorithm}

\section{Evaluation} \label{evalaute}

 To evaluate the effect of the dynamic wireless charging of EVs, we conducted simulations in the city of Erlangen.\par

\subsection{Evaluation Setup}

\begin{figure}
\begin{center}
\includegraphics[width=0.45\textwidth]{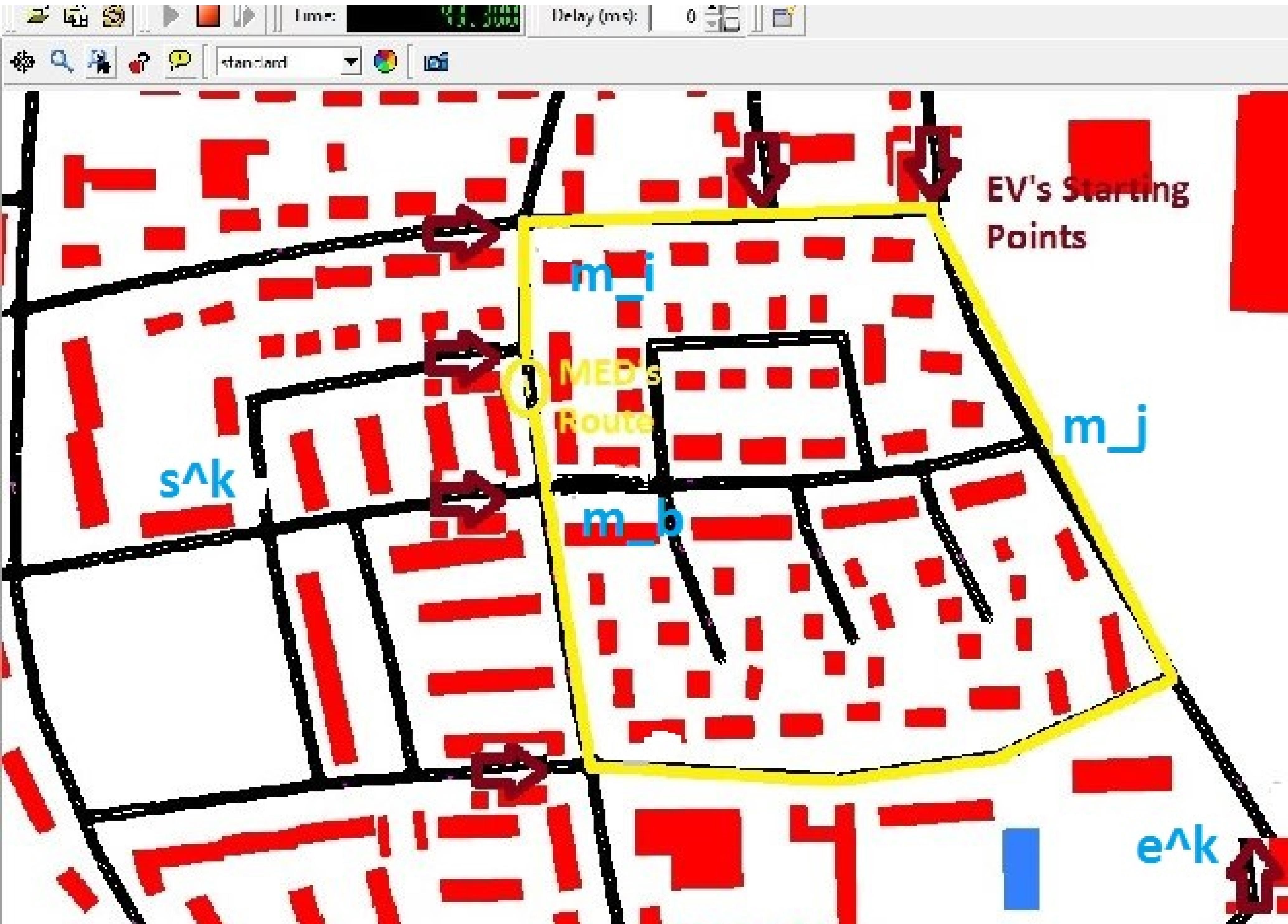}
\caption {Part of the Erlangen city map used for conducting the simulations. The MED route is marked in yellow. The position of the SCS is marked in green. The brown arrows are pointing at the starting points of the journeys of the EVs for both the SCS and the SCS + MED charging system.}
\label{fig-med-map}
\end{center}
\end{figure}

As can be seen in Figure \ref{fig-med-map} a bus which follows a specific route (shown in yellow in the figure) is used as a MED. On the other hand, a static charging station (SCS) is located at a fixed point at the road side of the corresponding city district. All the parametric side roads of the area in which the SCS and MED charging models are located are used as starting points $(s^k)$ for the dynamic wireless charging system with the same probability. The point at which the EVs are introduced in SCS or MED system is shown in Figure \ref{fig-med-map} with ($m_b,s_b$ respectively). The number of EVs that are inserted in the system is between 0 and 100. In addition, each EV $k$
entering the system has starting energy $\epsilon^k_s$ according to a uniform distribution with values between $1-6 kWh$. 

The only communication paths available are via the ad-hoc network and there is no other communication infrastructure. All the above parameters were not in favour of any charging method (MED or SCS). So, our evaluation location is assumed as a quite fairly area for our evaluation experiments.
The power of the antenna is $Ptx= 18dBm$ and the communication frequency $f$ is $5.9 Ghz$. In our simulations, we use a minimum sensitivity (Pth) of $-69 dBm$ to $-85 db$, which gives a transmission range of $130$ to $300$ meters, as can be seen in  Table \ref{arr:param}. As a result of the above transmission range, there is no communication with a few EVs. So, a number of EVs are excluded from the charging procedure because of the communication lost among EVs. This happens when the Signal-to-Interference-Ratio (SINR) threshold is below the 10 dB due to attenuation that is caused by the building obstacles of the city.   

\begin{center}
\begin{table}
\centering
\caption{Evaluation parameters}
\label{arr:param}
 \begin{tabular}{||c c||} 
 \hline
 Independent parameters & Range of values \\ [0.5ex]
 \hline\hline
 Number of vehicles & 0-100  \\ 
 \hline
 Initial Energy ($\epsilon^k_s$) & 1-6 kWh  \\
 \hline
 $C_{ind}$ & 0.7-0.8  \\
 \hline
 $P_{ind}$ & 20-50 kW  \\
 \hline
  $P_{tx}$ & 18dBm  \\
  \hline
  $f$ & 5.9Ghz    \\
  \hline
  Minimum sensitivity $(P_{th})$ & -69dBm to -85dB \\
  \hline
  Transmission range & $ 130-300$ meters\\
  \hline
 $n$ & 0.7-0.8  \\ 
 \hline
\end{tabular}
\end{table}
\end{center}

\subsection{Implementation of the Dynamic Charging System}
As described in Section \ref{communication}, all the EVs are informed for the waiting time ($wt_i$) either at the SCS $i$ or MED $i$ through the periodical communication with MEDs or SCS (using the CAM messages). As an example, assume that the EV $k$ is located at point ($s^k$) as starting point in Figure \ref{fig-med-map}. In order this EV to decide the best point for the insertion of dynamic charging system the Dijkstra’s algorithm is used (i.e. the Algorithm \ref{alg1}). The point $(m_b)$ is the best point for the MED system, while the point$(s_b)$ is the best point for insertion for the SCS system (see Figure \ref{fig-med-map}).\par

The value for minimization with our dynamic charging algorithm is the travel time for a vehicle between the starting point ($s^k$) and the target point ($e^k$). The total travel time if the vehicle chooses the SCS choice depends on the travel time between the ($s^k$,$s_b$) points, the charging time at the SCS, the waiting time here and the travel time between the points ($s_b$,$e^k$), for which the Dijkstra's algorithm is used again. The charging rate level of the EVs at the SCS is about $19,2 KW/sec$ ~\cite{battery_charger}. The waiting time at the SCS depends on the queue of the SCS and the driving time between ($s^k$,$s_b$) points. The vehicle periodically informed by the SCS about the current queue and all the bookings that SCS already has (Queue (Waiting) time ($wt_b$)). Based on its current distance to SCS and mean velocity it can compute the time that it will arrive to the SCS (Driving time ($dt$)). So it can compute the waiting time as: $WaitingTime= wt_b- dt(s^k,s_b)$.\par 

If the vehicle chooses the MED for its re-charging needs, travel time will be adjusted to reflect the travel time between the points ($s^k$,$m_b$), the waiting time of the vehicle at point $m_b$, the time at which the vehicle follows the MED and thus is charged (the vehicle $k$ follows the MED for the roads ($i$,$j$) which are defined from the binary variable $y_{ij}^k$) and the travel time from the last point ($m_j$) of the last road ($i$,$j$) in which the vehicle $m$ follows the MED to the destination point $(e^k)$ with the usage of Dijkstra's algorithm. At the starting point $(s^k)$, the vehicle at short intervals informed by the MED about its current position ($m_i$) and the booked road segments that MED already has. 

The electric vehicle also computes the closest point $(m_b)$ to meet the MED based on the MED's cycle and the vehicle's current position and the driving time, using mean velocity. Based on the charging coefficient the vehicle computes for how many road segments it needs to follow the MED and that way it can find the ending point ($m_j$). Based on the booking of the MED, its current position and meeting point $m_b$, the vehicle computes the waiting time (the time that it will need to wait for the MED to come free of any booking at meeting point $(m_b$)). If road segments $(m_b,m_j)$ are not booked then the waiting time will be : \\
$wt= dtMED(m_i,m_b)- dtVehicle(s^k,m_b)$. \\
If the above equation is negative then the vehicle will have to go for the next cycle of the MED: $wt= dtMED(m_i,m_b)+dtMED(m_b,m_b)-dtVehicle(s^k,m_b)$. If any road segments between $(m_b,m_j)$ are booked then the vehicle will have to go for the next cycle of the MED again. We must add that there is no upper limit on the waiting time of a vehicle until the MED will be available.\par
 
For the charging time of the vehicle from the MED, when the vehicle books the MED then it knows the point $(m_i)$, so it can compute the charging time based on mean velocity and the ending point $(m_j)$ of charging. In order to calculate the energy will be needed for each vehicle, the power consumption for each road traveled must be computed ~\cite{energy_consumption}. The energy cost of every road segment can be expressed as a proportion of the mean velocity. The velocity is the quotient of the distance of the road segment and the time that the vehicle will need to spend on this segment $(i,j)$, i.e. $T_{i,j}$, on average. The two forces that oppose the motion of an automobile are rolling friction, $F_{roll}$ and air resistance,$F_{air}$ ( ~\cite{dynamic_charging}).

\begin{equation}
\label{f_roll}
F_{roll}= \mu_{\varsigma}*m*g, F_{air}= \frac{1}{2}A * C * p * u^{2}
\end{equation}

where, $m$ is the mass of the car in $Kg$, $g = 9.8m/s^{2}$, $u$ is
the mean velocity in $m/s$ and $\mu_{\varsigma}$ is the rolling resistance
coefficient. $C$ is a dimensionless constant called the drag
coefficient that depends on the shape of the moving body, $A$ is
the silhouette area of the car $(m^{2})$ and $p$ is the density of the
air (about $1.2 kg/m^{3}$ at sea level at ordinary temperatures).
Typical values of $C$ for cars range from 0.35 to 0.50.
In constant-speed driving on a level road, the sum of $F_{roll}$
and $F_{air}$ must be just balanced by the forward force supplied
by the drive wheels. The power that a vehicle needs when
traveling with a steady speed is given by Equation \eqref{power}.
\begin{equation}
\label{power}
P=n*F_{Forward}*u=n (F_{roll}+F_{air})*u
\end{equation}
where, $n$ is the efficiency factor of the system. The energy cost of vehicle $k$ for traveling in road segment $(i,j)$ in kwh, i.e. $c_{ij}$, is calculated by Equation \eqref{energy_consumption}.
\begin{equation}
\label{energy_consumption}
c_{ij}= P* T_{ij}
\end{equation}
If the road segment belongs to the path of a MED, then the vehicle can increase its energy by induction. The amount of the induced energy is proportional to the total time that the
EV and the MED will stay connected. This time depends on the meeting point $(m_b)$ between the vehicle and the MED in relation to the total road segment length and the availability
of the MED. In order to represent the induced energy per hour to the EV, Equation \eqref{energy_consumption} is rewritten:
\begin{equation}
\label{induced_energy}
c_{ij}= P* T_{ij}- \rho
\end{equation}
In Equation \eqref{induced_energy} the $\rho$ is the induced energy to the vehicle $k$ and is given by:
\begin{equation}
\rho= t_{cont}* C_{ind}* P_{ind}
\end{equation}
$C_{ind}$ is the induction coefficient and $t_{cont}$ the time of contact between the MED and the EV. $P_{ind}$ is the power of the MED. The values of the above parameters can be seen in Table \ref{arr:param}. We ignore acceleration and deceleration phenomena.

\subsection{Starting Energy vs Power Consumption levels}
\begin{figure*}
\centering
\begin{subfigure}{.5\textwidth}
  \includegraphics[width=.8\linewidth]{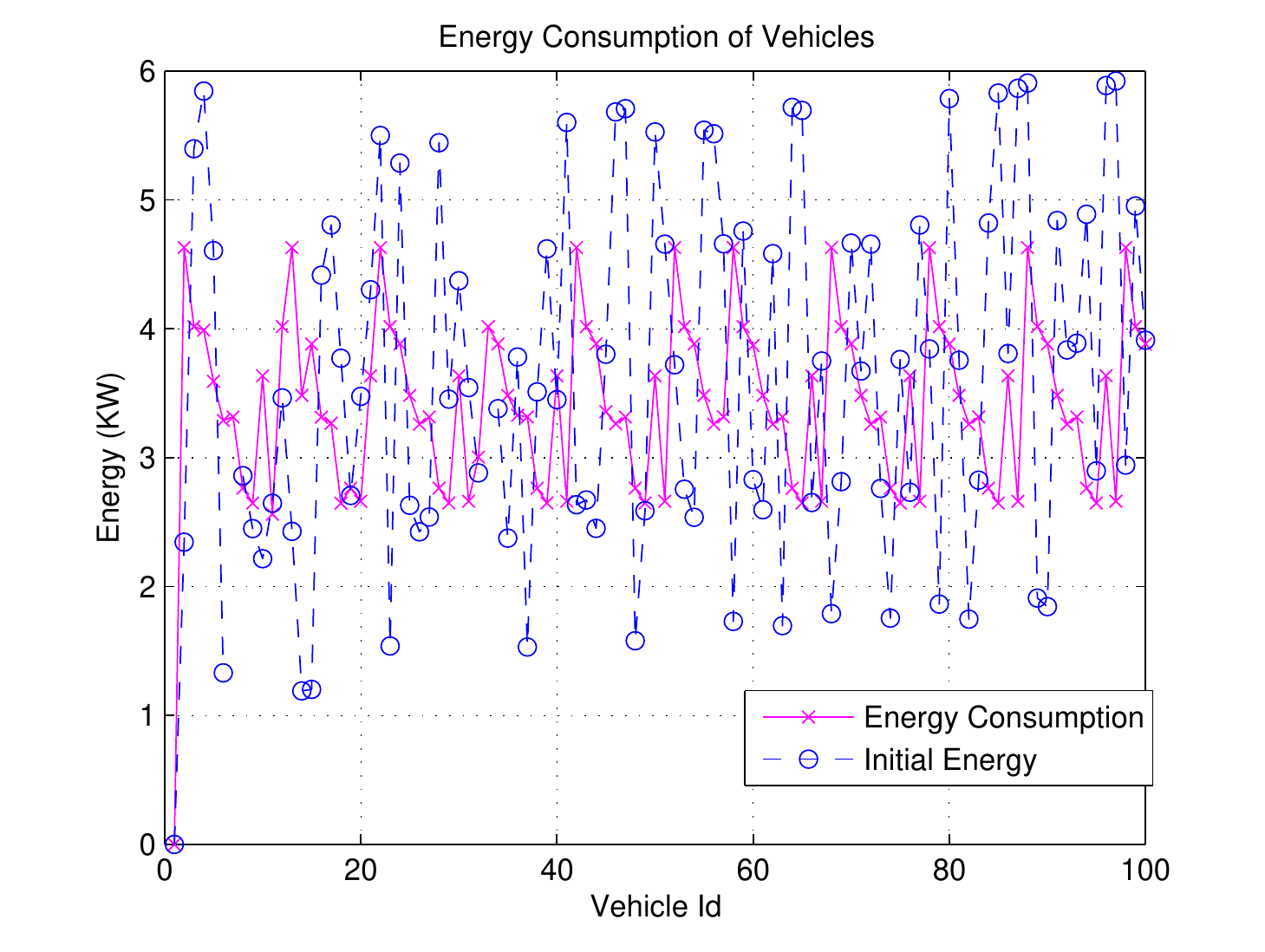}
  \caption{(LEVEL 1): The $20\%$ of EVs need re-charging}
  \label{fig:sub1}
\end{subfigure}%
~
\begin{subfigure}{.5\textwidth}
  \centering
  \includegraphics[width=.98\linewidth]{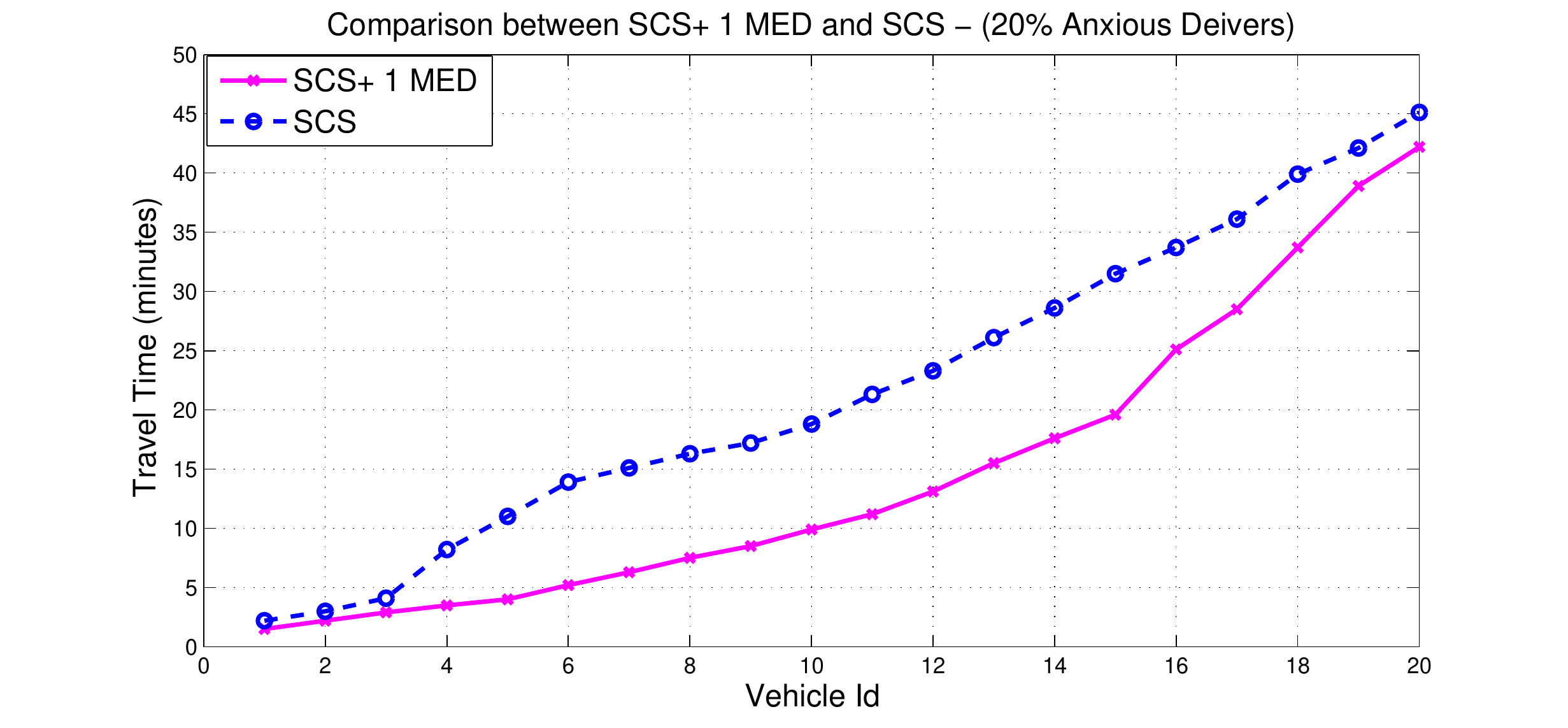}
  \caption{Travel Time when the $20\%$ of EVs need re-charging }
 \label{fig:res1}
\end{subfigure}
\begin{subfigure}{.5\textwidth}
\includegraphics[width=.8\linewidth]{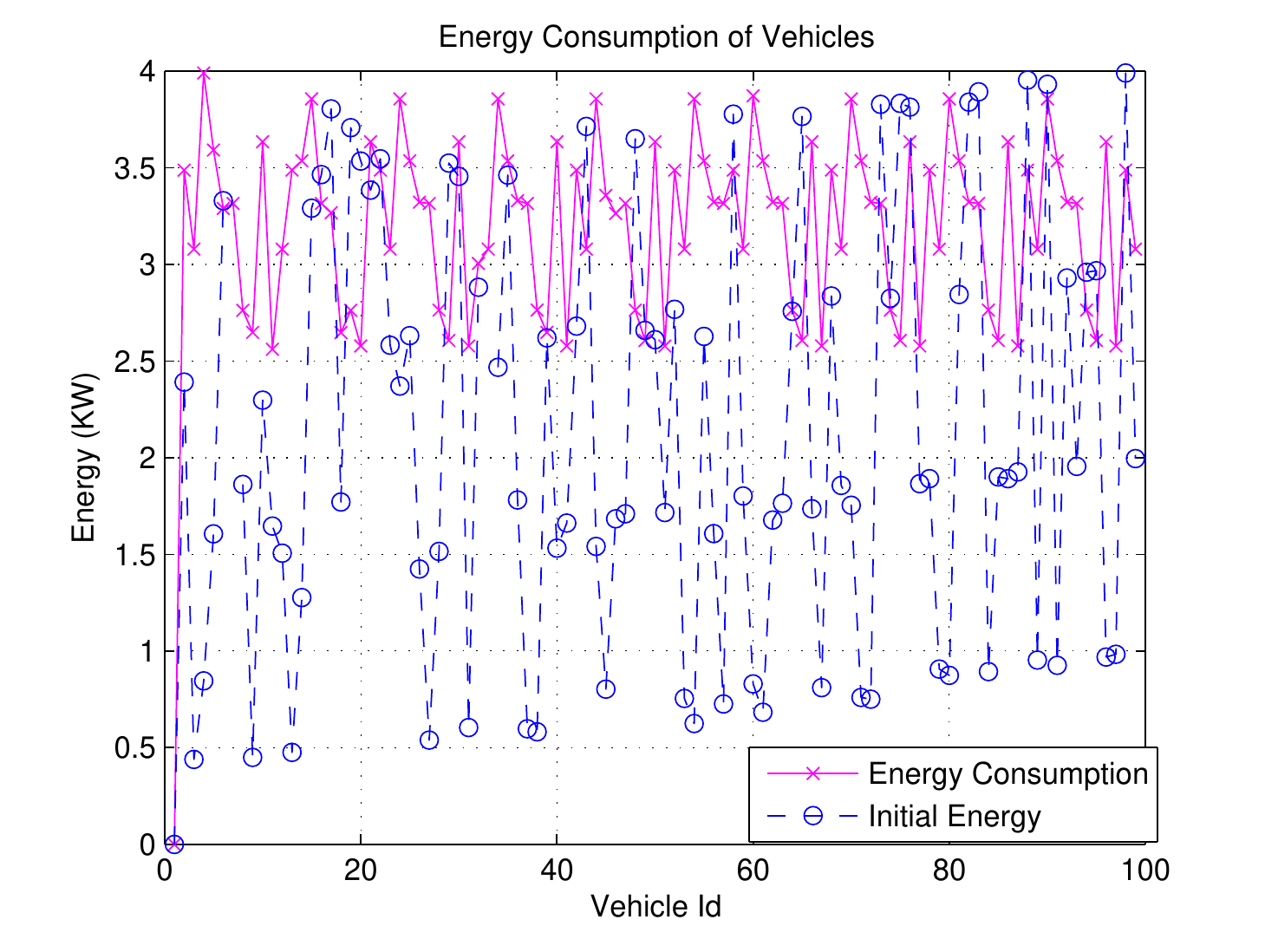}
\caption{(LEVEL 2): The $60\%$ of EVs need re-charging}
\label{fig:sub2}
\end{subfigure}%
~
\begin{subfigure}{.5\textwidth}
  \centering
\includegraphics[width=.98\linewidth]{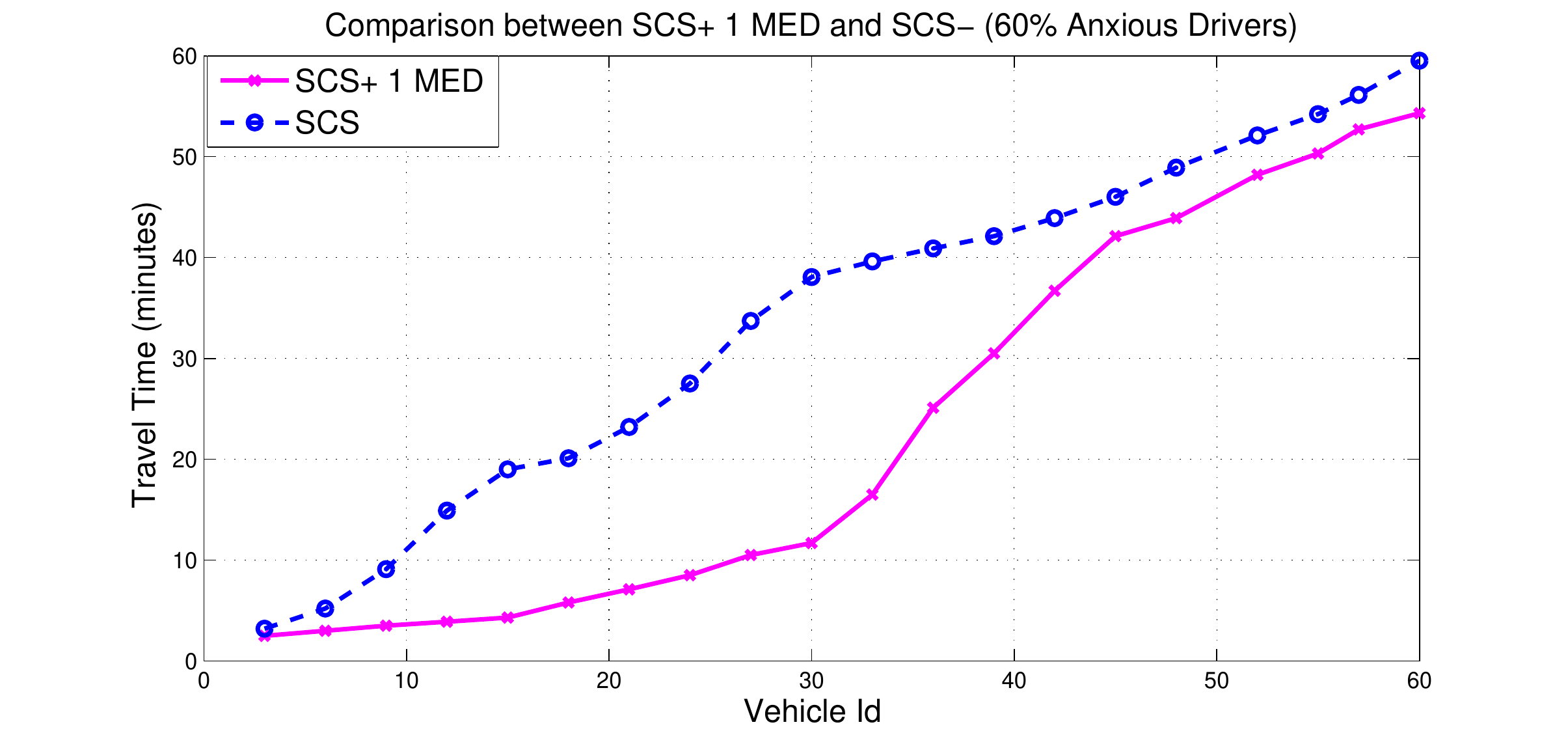}
\caption{Travel Time when the $60\%$ of EVs need re-charging }
\label{fig:res2}
\end{subfigure}
\begin{subfigure}{.5\textwidth}
 \includegraphics[width=.8\linewidth]{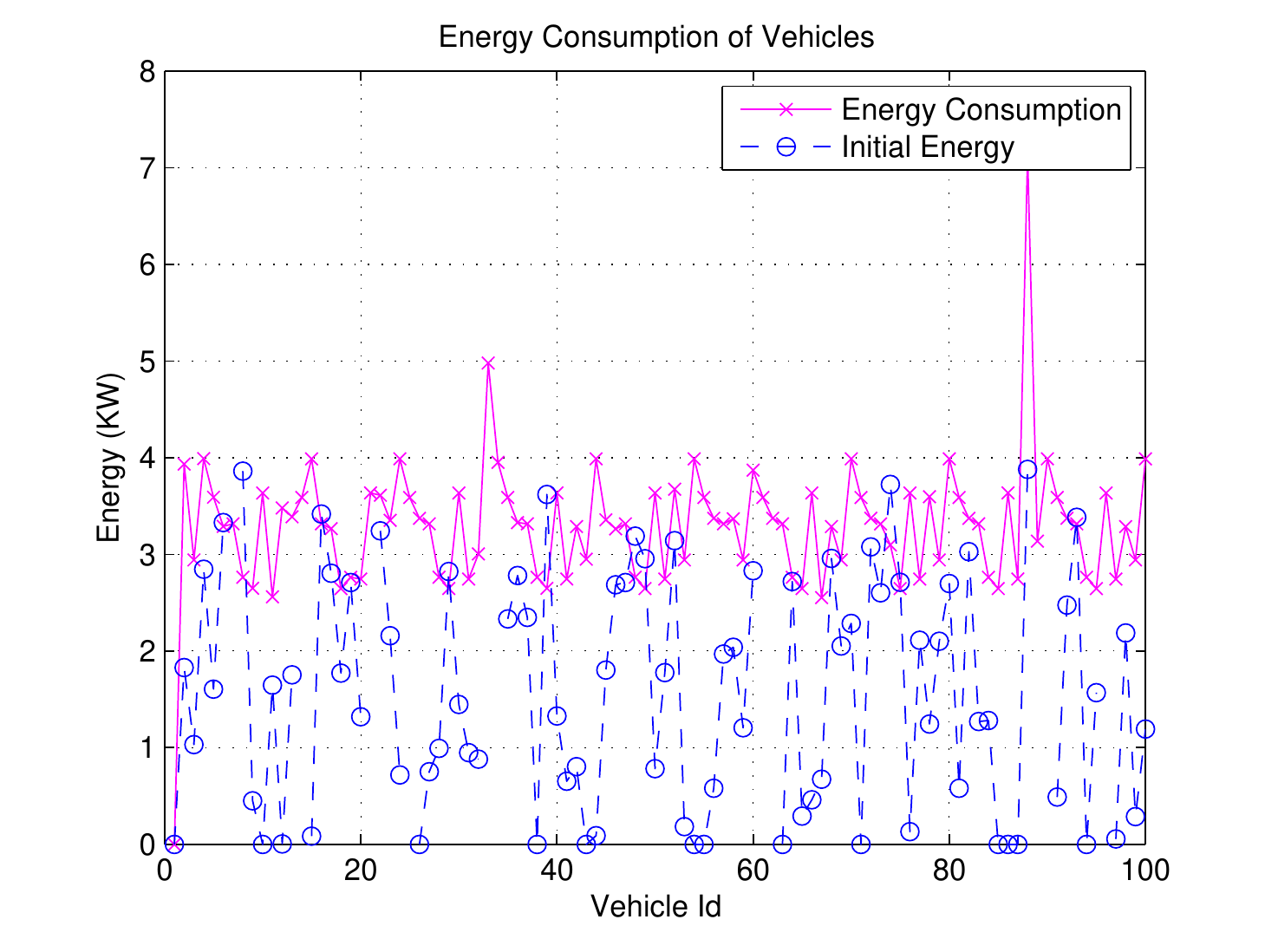}
 \caption{(LEVEL 3): The $95\%$ of EVs need re-charging}
 \label{fig:sub3}
\end{subfigure}%
~
\begin{subfigure}{.5\textwidth}
  \centering
 \includegraphics[width=.98\linewidth]{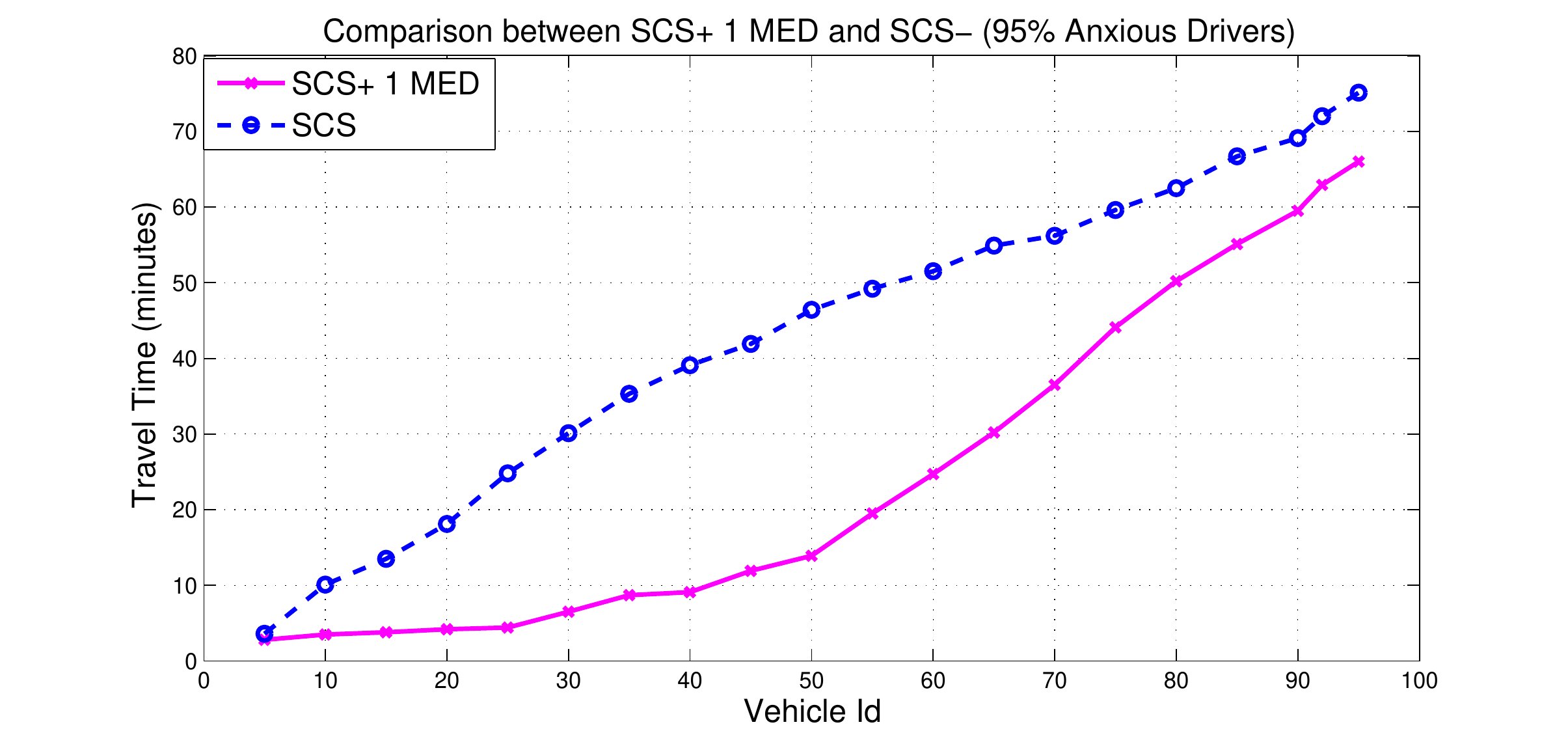}
 \caption{Travel Time when the $95\%$ of EVs need re-charging }
\label{fig:res3}
\end{subfigure}
\caption{Travel Time of all the Levels of Energy re-charging}
\end{figure*}

In our simulations, we used 3 levels of starting energy for the sum of the EVs. The starting energy for each EV is the remaining energy with which they approach the starting points of the system. We consider 3 different levels of the power consumption energy for the EVs in comparison with their initial energy. At the first level of the re-charging energy we consider that only the $20\%$ of EVs need re-charging in oder to reach at their destination (see Figure \ref{fig:sub1}). The second level of the power needs of the EVs is that in Figure \ref{fig:sub2}. Here $60\%$ of EVs need re-charging, increasing the complexity of the system. Last, at the third level of power need and initial energy comparison almost all the EVs need re-charging (the $95\%$ of EVs), as can be seen in Figure \ref{fig:sub3}. Contrary to the ~\cite{vtc2017} in which the number of drivers with range anxiety is a fixed number, this number is dynamic in our system and  depending on the EVs needs. All the drivers with initial energy smaller than the energy will be needed to be consumed are defined as anxious drivers.

\subsection{SCS vs. SCS + MED}
In this section we conduct a comparison of two different Charging System using 3 scenarios (see Figure (\ref{fig:sub1},\ref{fig:sub2},\ref{fig:sub3}). The first charging system contains only a static charging station, and the second charging system has a SCS and a MED. In Figures (\ref{fig:res1},\ref{fig:res2},\ref{fig:res3})  the travel time results for the above two system are presented. The sub-figures of Figure (\ref{fig:res1},\ref{fig:res2},\ref{fig:res3}) are corresponding to the charging needs of Figure (\ref{fig:sub1},\ref{fig:sub2},\ref{fig:sub3}).

Studying these results, it transpires that as the charging needs of vehicles are increasing, the travel time for both systems is also increasing. Specifically at the Level 1 of charging needs the travel time of the  dynamic charging model (SCS+ 1 MED) is better at about 2 times than that of the charging system (SCS) (see Figure \ref{fig:res1}), at the Level 2 the corresponding travel time of (SCS + 1 MED) is improved and is now at about 3 times better than that of the (SCS) (see Figure \ref{fig:res2}) and last at the Level 3 of re-charging needs the travel time is at about 4 times smaller than that of the only (SCS) model. 

Another observation from the results is that the travel time of the (SCS+MED) system is less than the (SCS) for all the circumstances of anxious drivers (0-100) and  energy charging need levels. For a small number of anxious drivers the difference between the two charging systems is very small. As the number of anxious drivers is increasing, the difference between the two systems is increasing too. However, when the number of anxious drivers is above  50 (for the Level 3 of charging needs) the difference is diminished. This behavior is due to the waiting time of the vehicles for the MED for a large number of cycles because of the preceding MED's bookings. 

Last, it is obvious that when the number of anxious drivers is above average of overall EVs the need of a MED in addition to a SCS is necessary, because the difference between the (SCS+ MED) system and the system that has only one SCS is bigger with $(60\%, 95\%)$ anxious drivers than that with $(20\%)$.

\subsection{SCS + MED system evaluation}
\begin{figure} 
\centering
\begin{subfigure}{.5\textwidth}
  \includegraphics[width=1.0\linewidth]{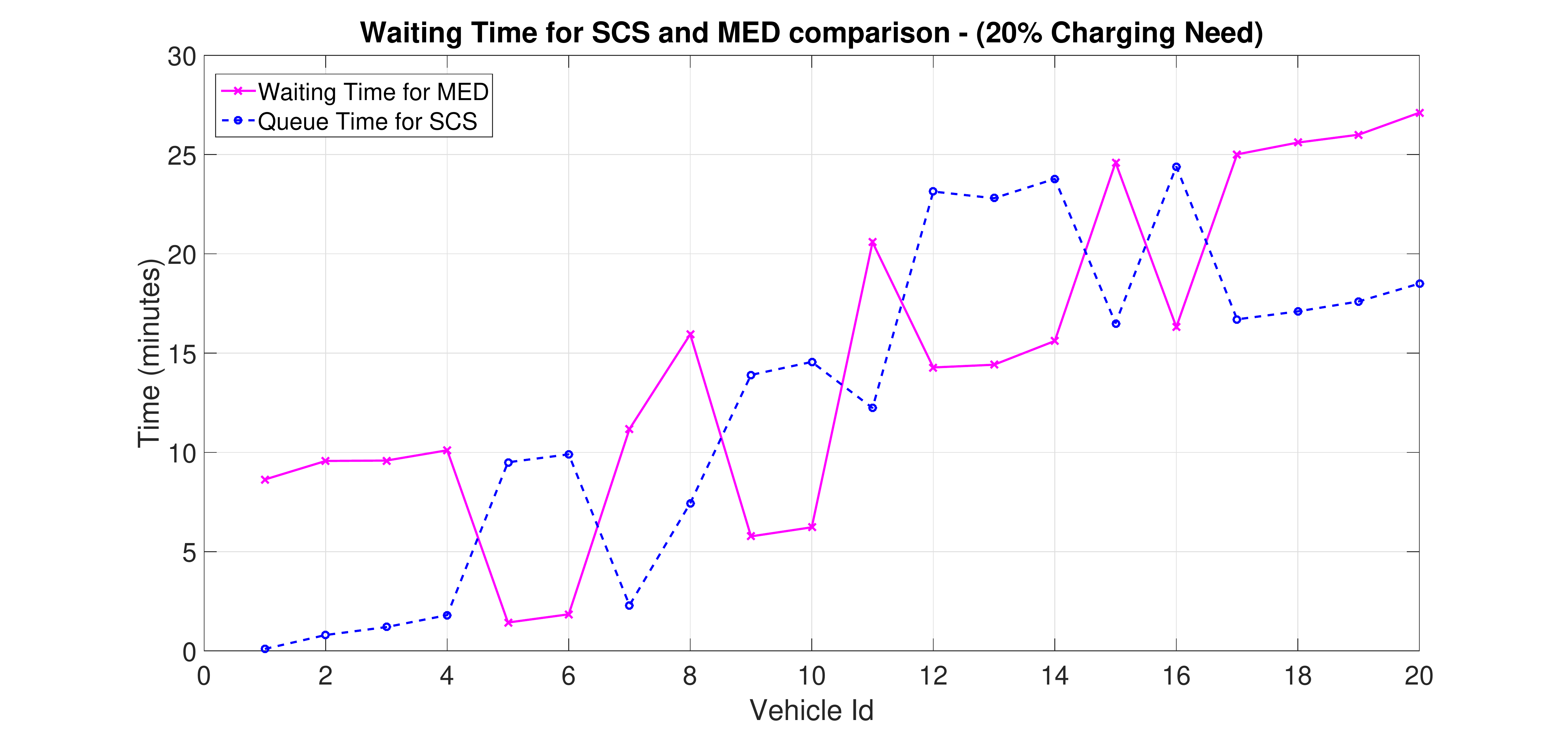}
  \caption{Waiting Time for MED vs Queue Time of SCS with the (LEVEL 1) of Energy re-charging}
  \label{fig:wt1}
\end{subfigure}%

\begin{subfigure}{.5\textwidth}
\includegraphics[width=1.0\linewidth]{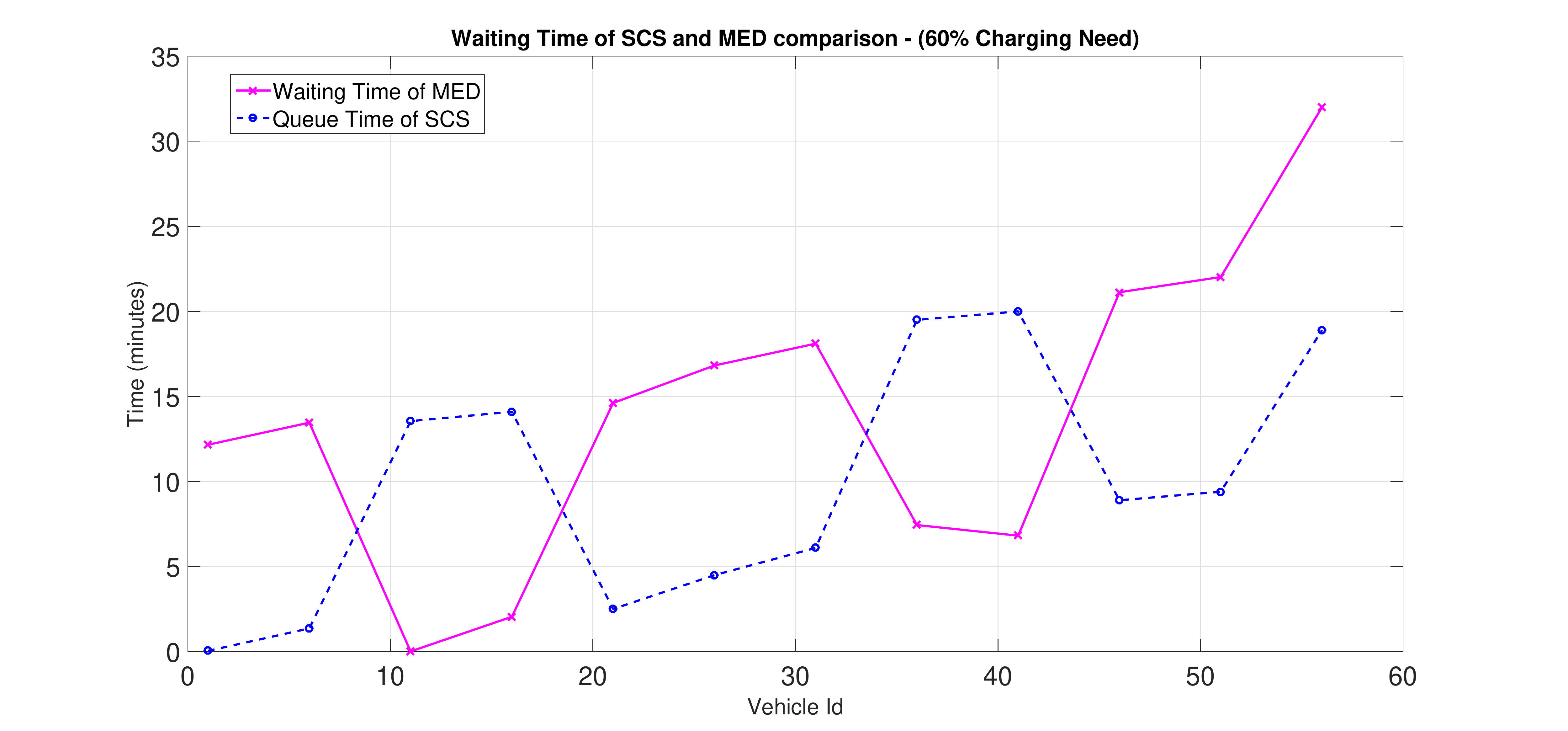}
\caption{Waiting Time for MED vs Queue Time of SCS with the (LEVEL 2) of Energy re-charging}
\label{fig:wt2}
\end{subfigure}%

\begin{subfigure}{.5\textwidth}
  \centering
 \includegraphics[width=1.0\linewidth]{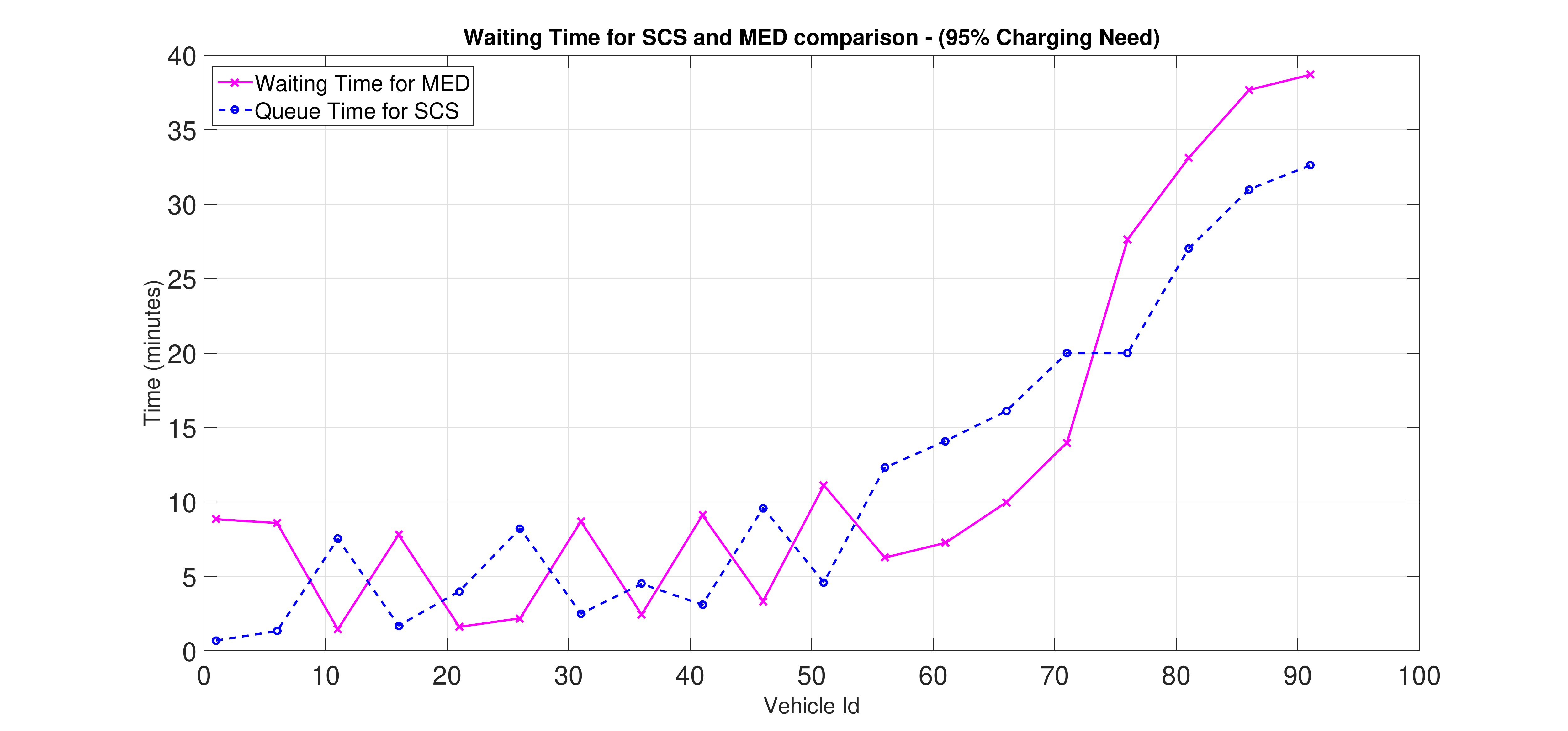}
 \caption{Waiting Time for MED vs Queue Time of SCS with the (LEVEL 3) of Energy re-charging}
\label{fig:wt3}
\end{subfigure}
\caption{Waiting Time for MED vs Queue Time of SCS}
\label{fig:wt}
\end{figure}

\begin{figure}
\centering
  \includegraphics[width=1.0\linewidth]{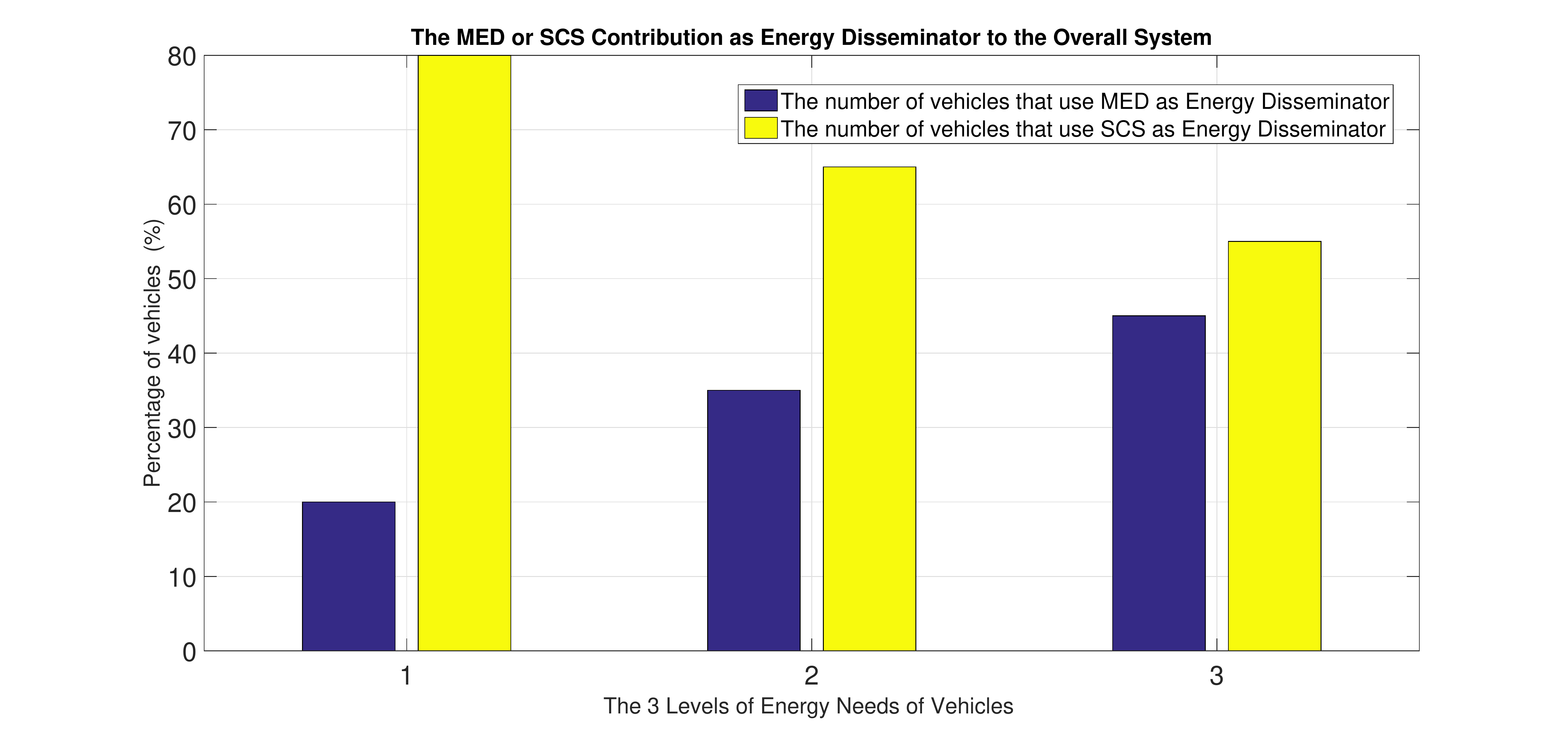}
  \caption{The number of EVs that select either the MED or the SCS for their re-charging needs for all the levels of Energy re-charging.}
  \label{fig:percentage}
\end{figure}

In this subsection the evaluation  of the system (SCS + 1 MED) is presented in more detail. In Figure \ref{fig:wt}, the waiting time $(wt)$ of each EV at the point $(m_b)$ that is planed to meet and follow the MED is compared with the Queue Time for each EV at the SCS. Moreover, in Figure \ref{fig:percentage} the percentage of EVs that select the MED or the SCS for re-charging is presented. We can see that as the number of anxious drivers is increasing, the number of EVs that select the MED as Energy Disseminator is increasing too. 

Studying more carefully Figure \ref{fig:wt}, it is obvious that at the starting time of the dynamic charging system when the queue of the SCS is empty and due to the fact that all EVs select the MED for re-charging results on the  increase of the waiting time. As the simulation time increases, the waiting time for MED and SCS are both widely fluctuated. This happens because the choice of EVs (MED or SCS) for re-charging are quickly interchanged. Studying the travel time of Figures (\ref{fig:sub1},\ref{fig:sub2},\ref{fig:res3}), a reduction of the difference of the travel time between the systems (SCS),(SCS + 1 MED) has observed. This phenomenon can be explained due to the increase of the waiting time at the MED, because of the frequent MED selection, (see Figure \ref{fig:wt}) when the number of anxious drivers increases (i.e. above 80 anxious drivers for Level 3 Energy re-charging). This leads  anxious drivers to choose the SCS and when its queue time increases, this situation reverses again.

Comparing the waiting time results for the 3 Levels of Energy re-charging, we can see that when the MED takes part more in EVs re-charging (see Figure \ref{fig:wt3}), the waiting time or queue time is not increased with such a steep mode as that of Figures (\ref{fig:wt1}, \ref{fig:wt2}). Moreover, the more interchanges between waiting time for MED or queue time for SCS in Figure \ref{fig:wt3} and the more usage of MED for EVs charging needs at the Level 3 (see Figure \ref{fig:percentage}) fully justify the wide difference between the two charging systems (SCS), (SCS +1 MED) in Figure \ref{fig:res3}.

\subsection{Cost Benefit Analysis}

There are several revenue possibilities stemming from this concept as well. Electric utilities, for example, might consider subsidizing the modification of trucks and buses into MEDs under a scenario in which the utility then becomes a revenue sharing partner with the MED owner. Governments at the state, local and national levels are all involved in policy-making decisions regarding environmental impact mitigation options, often using analytical tools \cite{Cassting16}. In this case, governments may consider offering tax incentives to modify trucks and buses into MEDs to further promote popularity and adoption of EVs.

In addition there are entrepreneurial advantages. Special software will need to be designed and refined for the physical platooning of the MEDs and EVs as well as for handling the appointment and billing logistics. Manufacturers will be asked to design and build the magnetic subsystems that create the foundation of the wireless charging systems. Converting a bus or truck into a MED would cost around $26,000$, while the cost of adding the technology to a passenger vehicle would be about $1,500$.

\section{Conclusions}\label{concl}

We have proposed a solution to increasing the driving range of electric vehicles based on modern communications between vehicles and state of the art technologies on energy transfer. The proposed solution steers away from larger and more powerful batteries, although these would still be useful and complements what we are proposing here. It does not require changes to existing road infrastructure which are costly and often pose health hazards. In contrast to vehicle-to-vehicle (V2V) charging schemes that are recently discussed in the literature \cite{v2v2012charging}, our work builds on the idea of using the city buses that follow predefined scheduled routes for dynamic charging in urban environments.

Combining modern communications between vehicles and state of the art technologies on energy transfer, we have shown that vehicles can extend their travel range. Energy exchange between vehicles can be facilitated by a process called “inductive power transfer” (IPT). This allows for an efficient and real-time energy exchange where vehicles can play an active role in the process.

Making use of inductive charging MEDs that act as mobile charging stations can improve the overall travel time of a fleet of vehicles compared to using only static charging stations. Specifically, using a MED in support of a SCS the overall travel time can be improved about four times compared with the only SCS usage case. The improvement of travel time comes with a negligible cost in travel distance, but starving vehicles otherwise would have to stop or make longer re-routes to find a stationary station and recharge their batteries.

As part of our future work, we intend to explore the evaluation of above dynamic charging method with a wide diversity of evaluation parameters. Specifically, we plan to use a bigger number of MEDs in combination with the existing SCSs and different areas of the city in order to further evaluate our dynamic charging system.

\balance
\bibliographystyle{IEEEtran}
\bibliography{main}

\begin{thebibliography}{10}
\providecommand{\url}[1]{#1}
\csname url@samestyle\endcsname
\providecommand{\newblock}{\relax}
\providecommand{\bibinfo}[2]{#2}
\providecommand{\BIBentrySTDinterwordspacing}{\spaceskip=0pt\relax}
\providecommand{\BIBentryALTinterwordstretchfactor}{4}
\providecommand{\BIBentryALTinterwordspacing}{\spaceskip=\fontdimen2\font plus
\BIBentryALTinterwordstretchfactor\fontdimen3\font minus
  \fontdimen4\font\relax}
\providecommand{\BIBforeignlanguage}[2]{{%
\expandafter\ifx\csname l@#1\endcsname\relax
\typeout{** WARNING: IEEEtran.bst: No hyphenation pattern has been}%
\typeout{** loaded for the language `#1'. Using the pattern for}%
\typeout{** the default language instead.}%
\else
\language=\csname l@#1\endcsname
\fi
#2}}
\providecommand{\BIBdecl}{\relax}
\BIBdecl

\bibitem{vegni2013smart}
A.~M. Vegni, M.~Biagi, and R.~Cusani, ``Smart vehicles, technologies and main
  applications in vehicular ad hoc networks,'' in \emph{Vehicular
  Technologies-Deployment and Applications}.\hskip 1em plus 0.5em minus
  0.4em\relax InTech, 2013.

\bibitem{thiel2015electric}
C.~Thiel, J.~Krause, and P.~Dilara, ``Electric vehicles in the eu from 2010 to
  2014-is full scale commercialisation near,'' \emph{Luxembourg: Publications
  Office}, 2015.

\bibitem{vtc2017}
E.~Bulut and M.~C. Kisacikoglu, ``Mitigating range anxiety via
  vehicle-to-vehicle social charging system,'' \emph{IEEE Vehicular Technology
  Conference}, 2017.

\bibitem{charging2015infrastructure}
A.~Hecker and R.~Wies, ``Charging infrastructure for evs in beijing: A spatial
  analysis from real customer data at two districts,'' \emph{Connected Vehicles
  and Expo (ICCVE), 2015 International Conference on}, pp. 336--341, 2015.

\bibitem{covic2013modern}
G.~A. Covic and J.~T. Boys, ``Modern trends in inductive power transfer for
  transportation applications,'' \emph{IEEE Journal of Emerging and Selected
  topics in power electronics}, vol.~1, no.~1, pp. 28--41, 2013.

\bibitem{plugless}
M.~Yamauchi,
  https://www.pluglesspower.com/learn/mainstream-electric-cars-are-headed-towards-wireless-charging/,
  2017.

\bibitem{maglaras2014cooperative}
L.~Maglaras, F.~V. Topalis, and A.~L. Maglaras, ``Cooperative approaches for
  dymanic wireless charging of electric vehicles in a smart city,'' in
  \emph{Energy Conference (ENERGYCON), 2014 IEEE International}.\hskip 1em plus
  0.5em minus 0.4em\relax IEEE, 2014.

\bibitem{maglaras2015-IJACSA}
L.~Maglaras, J.~Jiang, A.~Maglaras, F.~Topalis, and S.~Moschoyiannis, ``Dynamic
  wireless charging of electric vehicles on the move with mobile energy
  disseminators,'' \emph{Int'l Journal of Advanced Computer Science and
  Applications}, vol.~6, pp. 239--251, 2015.

\bibitem{jung2012high}
G.~Jung, B.~Song, S.~Shin, S.~Lee, J.~Shin, Y.~Kim, and S.~Jeon, ``High
  efficient inductive power supply and pickup system for on-line electric
  bus,'' in \emph{Electric Vehicle Conference (IEVC), 2012 IEEE
  International}.\hskip 1em plus 0.5em minus 0.4em\relax IEEE, 2012, pp. 1--5.

\bibitem{telewatt_2013}
A.~R. Mario, A.~A. Fadi, M.~Gagnaire, and Y.~Lascaux, ``Telewatt: An innovative
  electric vehicle charging infrastructure over public lighting systems,'' in
  \emph{Proceedings of the 2nd International Conference on Connected Vehicles
  and Expo (ICCVE), Las Vegas, USA, December}, 2013.

\bibitem{greencar}
``Innovative on street ev charging solutions,'' in \emph{Ecolane Consultancy
  and Next Green Car, White Paper}.\hskip 1em plus 0.5em minus 0.4em\relax
  Available:
  http://www.ecolane.co.uk/wp-content/uploads/2015/01/Ecolane-Innovative-on-street-EV-charging-solutions.pdf,
  2015.

\bibitem{xu2013optimal}
H.~Xu, S.~Miao, C.~Zhang, and D.~Shi, ``Optimal placement of charging
  infrastructures for large-scale integration of pure electric vehicles into
  grid,'' \emph{International Journal of Electrical Power \& Energy Systems},
  vol.~53, pp. 159--165, 2013.

\bibitem{adler2014online}
J.~D. Adler and P.~B. Mirchandani, ``Online routing and battery reservations
  for electric vehicles with swappable batteries,'' \emph{Transportation
  Research Part B: Methodological}, vol.~70, pp. 285--302, 2014.

\bibitem{de2016intention}
M.~M. de~Weerdt, S.~Stein, E.~H. Gerding, V.~Robu, and N.~R. Jennings,
  ``Intention-aware routing of electric vehicles,'' \emph{IEEE Transactions on
  Intelligent Transportation Systems}, vol.~17, no.~5, pp. 1472--1482, 2016.

\bibitem{lukic2013cutting}
S.~Lukic and Z.~Pantic, ``Cutting the cord: Static and dynamic inductive
  wireless charging of electric vehicles,'' \emph{IEEE Electrification
  Magazine}, vol.~1, no.~1, pp. 57--64, 2013.

\bibitem{ning2013compact}
P.~Ning, J.~M. Miller, O.~C. Onar, C.~P. White, and L.~D. Marlino, ``A compact
  wireless charging system development,'' in \emph{2013 Twenty-Eighth Annual
  IEEE Applied Power Electronics Conference and Exposition (APEC)}, 2013.

\bibitem{miller2014demonstrating}
J.~M. Miller, O.~C. Onar, C.~White, S.~Campbell, C.~Coomer, L.~Seiber, R.~Sepe,
  and A.~Steyerl, ``Demonstrating dynamic wireless charging of an electric
  vehicle: The benefit of electrochemical capacitor smoothing,'' \emph{IEEE
  Power Electronics Magazine}, vol.~1, no.~1, pp. 12--24, 2014.

\bibitem{nagendra2014detection}
G.~R. Nagendra, L.~Chen, G.~A. Covic, and J.~T. Boys, ``Detection of evs on ipt
  highways,'' \emph{IEEE journal of emerging and selected topics in power
  electronics}, vol.~2, no.~3, pp. 584--597, 2014.

\bibitem{v2v2012charging}
M.~C. Kisacikoglu, A.~Bedir, B.~Ozpineci, and L.~M. Tolbert, ``Phevev charger
  technology assessment with an emphasis on v2g operation,'' \emph{Oak Ridge
  Nat. Lab., Tech. Rep. ORNL/TM-2010/221}, Mar. 2012.

\bibitem{machiels2013design}
N.~Machiels, N.~Leemput, F.~Geth, J.~Van~Roy, J.~Buscher, and J.~Driesen,
  ``Design criteria for electric vehicle fast charge infrastructure based on
  flemish mobility behavior,'' 2013.

\bibitem{5289760}
G.~Putrus, P.~Suwanapingkarl, D.~Johnston, E.~Bentley, and M.~Narayana,
  ``Impact of electric vehicles on power distribution networks,'' in
  \emph{Vehicle Power and Propulsion Conference, 2009. VPPC '09. IEEE}, 2009.

\bibitem{Guho_2013}
G.~Jung, B.~Song, S.~Shin, S.~Lee, J.~Shin, Y.~Kim, C.~Lee, and S.~Jung,
  ``Wireless charging system for on-line electric bus(oleb) with
  series-connected road-embedded segment,'' in \emph{Environment and Electrical
  Engineering (EEEIC), 2013 12th International Conference on}, 2013.

\bibitem{pasaoglu2012driving}
G.~Pasaoglu, D.~Fiorello, A.~Martino, G.~Scarcella, A.~Alemanno, A.~Zubaryeva,
  and C.~Thiel, ``Driving and parking patterns of european car drivers-a
  mobility survey,'' \emph{Luxembourg: European Commission Joint Research
  Centre}, 2012.

\bibitem{brinknews}
L.~Maglaras, ``How to charge your electric car 'on the fly',''
  www.brinknews.com/how-to-charge-your-electric-car-on-the-fly/, 2017.

\bibitem{rehman2016adaptive}
O.~Rehman, M.~Ould-Khaoua, and H.~Bourdoucen, ``An adaptive relay nodes
  selection scheme for multi-hop broadcast in vanets,'' \emph{Computer
  Communications}, vol.~87, pp. 76--90, 2016.

\bibitem{basaras2013detecting}
P.~Basaras, D.~Katsaros, and L.~Tassiulas, ``Detecting influential spreaders in
  complex, dynamic networks,'' \emph{Computer}, vol.~46, no.~4, pp. 0024--29,
  2013.

\bibitem{santini2017consensus}
S.~Santini, A.~Salvi, A.~S. Valente, A.~Pescap{\'e}, M.~Segata, and R.~L.
  Cigno, ``A consensus-based approach for platooning with intervehicular
  communications and its validation in realistic scenarios,'' \emph{IEEE
  Transactions on Vehicular Technology}, vol.~66, no.~3, pp. 1985--1999, 2017.

\bibitem{stark2004wireless}
J.~C. Stark, ``Wireless power transmission utilizing a phased array of tesla
  coils,'' Ph.D. dissertation, Massachusetts Institute of Technology, 2004.

\bibitem{ramboz1996machinable}
J.~D. Ramboz, ``Machinable rogowski coil, design, and calibration,'' \emph{IEEE
  Transactions on Instrumentation and Measurement}, vol.~45, no.~2, pp.
  511--515, 1996.

\bibitem{haq2010new}
I.~Haq, R.~Monfared, R.~Harrison, L.~Lee, and A.~West, ``A new vision for the
  automation systems engineering for automotive powertrain assembly,''
  \emph{International Journal of Computer Integrated Manufacturing}, vol.~23,
  no.~4, pp. 308--324, 2010.

\bibitem{Dijkstra1959}
E.~W. Dijkstra, ``A note on two problems in connexion with graphs,''
  \emph{Numerische Mathematik}, vol.~1, no.~1, pp. 269--271, Dec 1959.

\bibitem{battery_charger}
M.~Yilmaz, Krein, and P.~T., ``Review of battery charger topologies, charging
  power levels, and infrastructure for plug-in electric and hybrid vehicles,''
  \emph{IEEE Transactions on Power Electronics}, vol.~28, pp. 2151--2169, 2013.

\bibitem{energy_consumption}
X.~Wu, F.~David, K.~Alfredo~Cabrera, and W.~A., ``Electric vehicles’ energy
  consumption measurement and estimation,'' \emph{IEEE Vehicular Technology
  Conference}, vol.~34, pp. 52--67, 2015.

\bibitem{dynamic_charging}
L.~A. Maglaras, J.~Jiang, A.~Maglaras, F.~V. Topalis, and S.~Moschoyiannis,
  ``Dynamic wireless charging of electric vehicles on the move with mobile
  energy disseminators,'' \emph{International Journal of Advanced Computer
  Science and Applications(ijacsa)}, vol.~6, 2015.

\bibitem{Cassting16}
S.~Moschoyiannis, N.~Elia, A.~Penn, D.~J.~B. Lloyd, and C.~Knight, ``A
  web-based tool for identifying strategic intervention points in complex
  systems,'' in \emph{Proc. Games for the Synthesis of Complex Systems
  (CASSTING'16 @ ETAPS 2016)}, ser. EPTCS, vol. 220, 2016, pp. 39--52.

\end{thebibliography}

\end{document}